\documentclass[12pt, a4paper]{article}
\usepackage[utf8]{inputenc}
\usepackage[left=2cm, right=2cm, bottom=2cm, top=2cm]{geometry}
\usepackage{bm}
\usepackage{amsmath, amsfonts}
\usepackage{url}
\usepackage{svg}
\usepackage{booktabs}
\usepackage{graphicx}
\usepackage{caption}
\usepackage{subcaption}
\usepackage{hyperref}
\usepackage[capitalise]{cleveref}
\usepackage{csvsimple}
\usepackage{authblk}
\usepackage{mathtools}
\usepackage{float}
\usepackage{xcolor}

\usepackage[square, numbers, comma, sort&compress]{natbib}

\title{Impact of population size on early adaptation in rugged fitness landscapes}
\author{Richard Servajean\textsuperscript{1,2}, Anne-Florence Bitbol\textsuperscript{1,2,*}}
\affil{\textbf{1} Institute of Bioengineering, School of Life Sciences, École Polytechnique Fédérale de Lausanne (EPFL), CH-1015 Lausanne, Switzerland\

\textbf{2} SIB Swiss Institute of Bioinformatics, CH-1015 Lausanne, Switzerland\
*Email: anne-florence.bitbol@epfl.ch}
\date{}

\begin{document}
\maketitle

\begin{abstract}
Due to stochastic fluctuations arising from finite population size, known as genetic drift, the ability of a population to explore a rugged fitness landscape depends on its size. In the weak mutation regime, while the mean steady-state fitness increases with population size, we find that the height of the first fitness peak encountered when starting from a random genotype displays various behaviors versus population size, even among small and simple rugged landscapes. We show that the accessibility of the different fitness peaks is key to determining whether this height overall increases or decreases with population size. Furthermore, there is often a finite population size that maximizes the height of the first fitness peak encountered when starting from a random genotype. This holds across various classes of model rugged landscapes with sparse peaks, and in some experimental and experimentally-inspired ones. Thus, early adaptation in rugged fitness landscapes can be more efficient and predictable for relatively small population sizes than in the large-size limit. 
\end{abstract}

\section{Introduction}

Natural selection drives populations towards higher fitness (i.e.\ reproductive success), but actual fitness landscapes (representing fitness versus genotype~\cite{wright1932,smith1970}) can possess several distinct local maxima or peaks. Such rugged fitness landscapes arise from epistasis, i.e.\ interactions between genetic variants~\cite{visser2011,poelwijk2011,visser2014,visser2018}, especially from reciprocal sign epistasis~\cite{poelwijk2011}, where two mutations together yield a benefit while they are deleterious separately, giving rise to a fitness valley~\cite{Dawid10,Draghi13}. While the high dimension of genotype space makes it challenging to probe fitness landscapes~\cite{Franke11,szendro2013b}, experimental evidence has been accumulating for frequent landscape ruggedness~\cite{Dawid10, Bloom10,Kryazhimskiy11b,szendro2013b,Draghi13,Covert13,bank2016,visser2018,Fragata19}. This strongly impacts the predictability of evolution~\cite{visser2014,bank2016,visser2018}. Populations can remain stuck at a local fitness peak, thus preventing further adaptation. Which local peak is reached depends on the starting point, on the mutations that occurred, on their order, and on whether they took over or not. Historical contingency may thus play important roles.

In a constant environment, if mutations are rare, the evolution of a homogeneous population of asexual microorganisms can be viewed as a biased random walk in genotype space, and thus on the associated fitness landscape~\cite{orr2005}. Indeed, random mutations can either fix (i.e.\ take over) or get extinct, depending on how mutant fitness compares to wild-type fitness (natural selection) and on stochastic fluctuations due to finite population size (genetic drift). In the weak mutation regime, mutations are rare enough for their fate to be sealed before a new mutation takes place. Thus, the population almost always has a single genotype, i.e.\ it is monomorphic. When a mutant fixes, it becomes the new wild type: the population has moved in genotype space -- hence the biased random walk in genotype space. If in addition natural selection is strong~\cite{orr2005,gillespie1983}, only beneficial mutations, which increase fitness, can fix. In this regime, the random walk describing the evolution of the population can only go upwards in fitness. Such \textit{adaptive walks} (AWs)~\cite{orr2005} have been extensively studied~\cite{nowak2015}. Strong selection neglects the possibility that deleterious or neutral mutations may fix due to genetic drift, which is appropriate only for very large populations~\cite{hartl1998}. Conversely, if the strong selection hypothesis is dropped, deleterious mutations may fix~\cite{mccandlish2011,mccandlish2014}, and a population's ability to explore its fitness landscape depends on its size, which determines the amplitude of genetic drift~\cite{ewens1979}. How does the interplay between genetic drift and natural selection~\cite{blount2018} impact adaptation of a finite-size population on rugged fitness landscapes? In particular, is adaptation always more efficient for larger populations?

To address this question, we consider homogeneous populations of constant size $N$, which evolve in the weak mutation regime, either through the Moran model~\cite{moran1958,ewens1979}, or through the Wright-Fisher model under the diffusion approximation~\cite{Kimura1962,crow2009}. The steady-state properties of such evolution have been studied, in particular the stationary distribution of states~\cite{Berg04,sella2005} and their dynamical neighborhoods~\cite{mccandlish2018}. The mean steady-state fitness monotonically increases with population size (see Supplementary material, \cref{sec:steadyfit} and \cref{fig:fitness_vs_time}), so the long-term outcome of evolution becomes more optimal and predictable when population size increases. A very large finite population will reach the highest fitness maximum of the landscape, but this may take very long, due to the difficulty of crossing fitness valleys for large populations with rare mutations. Here, we investigate the dynamics of adaptation before steady state is reached, and we ask how population size impacts early adaptation.

We focus on early adaptation, by considering the first fitness peak  encountered starting from a randomly chosen genotype. We mainly study the fitness of this first encountered peak, and we also discuss the time needed to reach it. Both have been extensively studied for adaptive walks~\cite{nowak2015}. We find that, in contrast to the steady-state fitness, the fitness $\bar{h}$ of the first encountered peak, averaged over starting genotypes, does not always increase with population size $N$. Thus, adaptation is not always more efficient for larger populations. Furthermore, we observe a wide variety of behaviors of $\bar{h}$ with $N$, even among small and simple rugged landscapes. We show that the accessibility of the different fitness peaks is a key ingredient to determine whether $\bar{h}$ is larger or smaller for large $N$ than for $N=1$. We find that the ensemble mean $\left\langle\bar{h}\right\rangle$ of $\bar{h}$ over different model landscapes often features a maximum for a finite value of $N$, showing that early adaptation is often most efficient for intermediate $N$. This effect occurs in rugged landscapes with low densities of peaks, is particularly important for large genomes with pairwise epistasis, and matters for larger populations when genomes are large. More generally, such finite-size effects extend to larger populations when many mutations are (almost) neutral. These situations are relevant in practice. Furthermore, our main conclusions hold for multiple experimental and experimentally-motivated landscapes.

\section{Methods}
\label{sec:model}

\subsection{Model}
\label{subsec:framework}

We consider a homogeneous population comprising a constant number $N$ of asexual haploid individuals, e.g. bacteria. We assume that their environment is constant, and we neglect interactions between genotypes (individual types, characterized by the state of all genes) and frequency-dependent selection. Each genotype is mapped to a fitness through a fitness landscape~\cite{wright1932,smith1970}, which is static under these hypotheses~\cite{visser2018}. We consider various rugged fitness landscapes (see Results).

Evolution is driven by random mutations, corresponding to a genotype change in one organism. The genotype of each organism is described by a sequence of $L$ binary variables, taking values $0$ or $1$, which correspond to nucleotides, amino acids, genes or any other relevant genetic unit. The binary state is a simplification~\cite{zagorski2016}, which can represent the most frequent state ($0$) and any variant ($1$). Genotype space is then a hypercube with $2^L$ nodes, each of them having $L$ neighbors accessible by a single mutation (i.e.\ a substitution from $0$ to $1$ or vice-versa at one site). Note that we do not model insertions or deletions. For simplicity, we further assume that all substitutions have the same probability. 

Because the population size $N$ is finite and there is no frequency-dependent selection, each mutant lineage either fixes (i.e.\ takes over the population) or gets extinct, excluding coexistence between quasi-stable clades~\cite{good2017}. We focus on the weak mutation regime, defined by $N\mu \ll 1$ where $\mu$ denotes mutation probability per site and per generation. Mutations are then rare enough for their fate to be sealed before any new mutation takes place. Thus, the population almost always has a single genotype, i.e.\ it is monomorphic, excluding phenomena such as clonal interference~\cite{elana2003,Laessig17,good2017}. When a mutant fixes, it becomes the new wild type. In this framework, the evolution of the population by random mutations, natural selection and genetic drift can be viewed as a biased random walk in genotype space~\cite{orr2005}. A mutation followed by fixation is one step of this random walk, where the population hops from one node to another in genotype space. To describe the fixation of a mutation in a homogeneous population of size $N$ under genetic drift and natural selection, we consider two population genetics models: the Moran model~\cite{moran1958,ewens1979} and the Wright-Fisher model under the diffusion approximation~\cite{Kimura1962,crow2009}, yielding two specific walks in genotype space. We use these models within the origin-fixation approach~\cite{mccandlish2014}, where the mutation fixation rate is written as the mutation origination rate times the fixation probability. Note that we assume that fitness is positive, as it represents division rate, requiring minor modifications for some fitness landscape models (\cref{subsec:overview}). 

\paragraph{Moran walk.} In the Moran process, at each step, an individual is picked to reproduce with a probability proportional to its fitness, and an individual is picked to die uniformly at random~\cite{moran1958,ewens1979}. The fixation probability of the lineage of one mutant individual with genotype $j$ and fitness $f_j$ in a wild-type population with genotype $i$ and fitness $f_i$ reads~\cite{ewens1979}:
\begin{align}\label{eq:moran}
\begin{split}
     P_{ij} &= \frac{1-\frac{f_{i}}{f_j}}{1-\left(\frac{f_{i}}{f_j}\right)^N} = \frac{1-(1+s_{ij})^{-1}}{1-(1+s_{ij})^{-N}} \text{ if } s_{ij} \neq 0\,,  \\
     P_{ij}  &= \frac{1}{N} \text{ if } s_{ij} = 0\,,
     \end{split}
 \end{align}   
where $s_{ij} = f_{j} / f_{i} - 1$. In the Moran walk, all mutations (substitutions at each site) are equally likely, and when a mutation arises, it fixes with probability $P_{ij}$. If it does, the population hops from node $i$ to node $j$ in genotype space. The Moran walk is a discrete Markov chain, where time is in number of mutation events, and it is irreducible, aperiodic and positive recurrent (and thus ergodic). Hence, it possesses a unique stationary distribution towards which it converges for any initial condition~\cite{norris,aldous2002}. It is also reversible~\cite{sella2005,mccandlish2018}. Note that evolution in the strong selection weak mutation regime (large-$N$ limit of the present case) yields absorbing Markov chains, with different properties.

\paragraph{Wright-Fisher walk.} The Wright-Fisher model assumes non-overlapping generations, where the next generation is drawn by binomial sampling~\cite{crow2009}. Under the diffusion approximation valid for large populations ($N \gg 1$) and mutations of small impact ($|s_{ij}|\ll 1$)~\cite{crow2009,Kimura1962}, the fixation probability of mutant $j$ reads
\begin{align}\label{eq:transition_proba_wf}
\begin{split}
    P_{ij} &= \frac{1-e^{-2s_{ij}}}{1-e^{-2Ns_{ij}}}\text{ if } s_{ij} \neq 0\,,  \\
     P_{ij} &= \frac{1}{N} \text{ if } s_{ij} = 0\,. 
\end{split}
\end{align}
We use this formula similarly as above to define the Wright-Fisher walk, which is also an irreducible, aperiodic, positive recurrent and reversible discrete Markov chain converging to a unique stationary distribution. Note that we use it for all $N$ and fitness values, but that it rigorously holds only under the diffusion approximation, i.e.\ under assumptions of large population and weak selection. In fact, the complete Wright-Fisher model is irreversible for large selection, although this does not impact the steady-state distribution of populations on fitness landscapes~\cite{manhart2012}. By contrast, the Moran fixation probability is exact within the Moran process.

\subsection{Quantifying early adaptation}
\label{subsec:obs}

To investigate early adaptation, and its dependence on population size, we mainly focus on the height $h$ of the walk, which is the fitness of the first encountered peak~\cite{nowak2015}. It depends on the initial node $i$, and also on what happens at each step of the walk. We consider the average $\bar{h}$ of $h$ over many walks and over all possible initial nodes, assumed to be equally likely: this quantity globally characterizes early adaptation in the fitness landscape. Starting from a random node is relevant e.g. after an environmental change which made the wild type no longer optimal, and allows to characterize early adaptation over the whole fitness landscape. We also study the impact of restricting the set of starting points to those with high fitness (see also~\cite{orr2005}), which is relevant for small to moderate environmental changes. To assess the variability of $h$, we also consider its standard deviation $\sigma_h$. Note that by definition, the height $h$ is directly the fitness value of a peak.

In addition to $\bar{h}$, we study the walk length $\bar{\ell}$, and its time $\bar{t}$, which are respectively the mean number of successful fixations and of mutation events (leading to fixation or not) before the first peak is reached, with similar methods as for $\bar{h}$ (see Supplementary material, \cref{sec:length_time}).

\paragraph{First step analysis (FSA).} To express $\bar{h}$, we consider the first hitting times of the different peaks (local fitness maxima) of the landscape~\cite{aldous2002}. Denoting by $M$ the set of all nodes that are local maxima and by $T_j$ the first hitting time of $j\in M$, we introduce the probability $P_i\left(T_j = \min \left[T_k, k \in M\right]\right)$ that a walk starting from node $i$ hits $j$ before any other peak. Discriminating over all possibilities for the first step of the walk (FSA) yields
\begin{equation}
  P_i\left(T_j = \min \left[T_k, k \in M\right]\right) =
    \begin{cases}
      1 & \text{if } i = j\,,\\
      0 & \text{if } i \in M \text{ and } i \neq j\,,\\
      \sum_{l \in G_i} \tilde{P}_{il} \,P_l\left(T_j = \min \left[T_k, k \in M\right]\right) & \text{otherwise}\,,
    \end{cases}   
\label{eq:systemfsa}
\end{equation}
where $G_i$ is the set of neighbors of $i$ (i.e.\ the $L$ genotypes that differ from $i$ by only one mutation), while $\tilde{P}_{il}=P_{il}/\sum_{q \in G_i}P_{iq}$, where $P_{il}$ is the fixation probability of the mutation from $i$ to $l$, given by \cref{eq:moran} or \cref{eq:transition_proba_wf}. Thus, $\tilde{P}_{il}$ is the probability to hop from $i$ to $l$ at the first step of the walk. Solving this system of $2^L n_M$ equations, where $n_M$ is the number of local maxima in the fitness landscape, yields all the first hitting probabilities. This allows to compute
\begin{equation}
    \bar{h} = \frac{1}{2^L}\sum_{i \in G} \bar{h}_i \,,\,\,\,\,\text{with}\,\,\,\,\bar{h}_i =\sum_{j \in M}f_j\, P_i\left(T_j = \min \left[T_k, k \in M\right]\right)\,,
\end{equation}
where $G$ is the ensemble of all the nodes of the landscape. Note that if the landscape has only two peaks $j$ and $k$, it is sufficient to compute $P_i(T_j<T_k)$ for all $i$, which can be expressed from the fundamental matrix of the irreducible, aperiodic, positive recurrent and reversible Markov chain corresponding to the Moran or Wright-Fisher walk~\cite{aldous2002,mccandlish2018}. These first hitting probabilities also allow us to compute the standard deviation $\sigma_h$ of $h$.

In practice, we solve \cref{eq:systemfsa} numerically using the NumPy function \texttt{linalg.solve}. Note however that since the number of equations increases exponentially with $L$ and linearly with $n_M$, this is not feasible for very large landscapes.

\paragraph{Stochastic simulations.} We also perform direct stochastic simulations of Moran and Wright-Fisher walks based on \cref{eq:moran} or \cref{eq:transition_proba_wf}, using a Monte Carlo procedure. Note that we simulate the embedded version of these Markov chains, where the transition probabilities to all neighbors of the current node are normalized to sum to one, avoiding rejected moves. The only exception is when we study the time $t$ of the walks, which requires including mutations that do not fix.

\paragraph{Averaging over multiple fitness landscapes.} To characterize adaptation in an ensemble of landscapes, we consider the ensemble mean $\left\langle \bar{h} \right\rangle$ of $\bar{h}$ by sampling multiple landscapes from the ensemble, and taking the average of $\bar{h}$, either computed by FSA or estimated by simulations.

\paragraph{Code availability.} \url{https://github.com/Bitbol-Lab/fitness-landscapes}

\section{Results}
\subsection{Early adaptation on $LK$ fitness landscapes}
\label{subsec:LK_model}

The $LK$ model (originally called $NK$ model) describes landscapes with tunable epistasis and ruggedness~\cite{kauffman1989}. In this model, the fitness of genotype $\vec{\sigma} = (\sigma_1, \sigma_2, ..., \sigma_L) \in \{0, 1\}^L$ reads
\begin{equation}
   f(\vec{\sigma}) = \sum_{i=1}^L f_i\left(\{\sigma_j\}_{j \in \nu_i}\right),
\label{eq:$LK$_model}
\end{equation}
where $f_i\left(\{\sigma_j\}_{j \in \nu_i}\right)$ denotes the fitness contribution associated to site $i$, and $\nu_i$ is the set of epistatic partners of $i$, plus $i$ itself. Here $L$ is genome length, i.e.\ the number of binary units (genes or nucleotides or amino acids) that characterize genotype, while $K$ is the number of epistatic partners of each site $i$ -- thus, for each $i$, $\nu_i$ comprises $K+1$ elements. Unless mentioned otherwise, we consider $LK$ landscapes where sets of partners are chosen uniformly at random, i.e.\ in a ``random neighborhood'' scheme~\cite{nowak2015}, and each fitness contribution $f_i\left(\{\sigma_j\}_{j \in \nu_i}\right)$ is independently drawn from a uniform distribution between 0 and 1. Epistasis increases with $K$. For $K=0$, all sites contribute additively to fitness. For $K=1$, each site $i$ has one epistatic partner, whose state impacts $f_i$. For $K>1$, there is higher-order epistasis. For $K=L-1$, all fitness contributions change when the state of one site changes, yielding a House of Cards landscape~\cite{Kauffman1987,kingman1978} where the fitnesses of different genotypes are uncorrelated.

How does finite population size $N$ impact the average height $\bar{h}$ of the first peak reached by an adapting population starting from a uniformly chosen genotype? We first tackle this question in $LK$ landscapes with $L = 3$ and $K = 1$, which are small and simple rugged landscapes. 

\begin{figure}[h!]
 \centering
 \includegraphics[width=\textwidth]{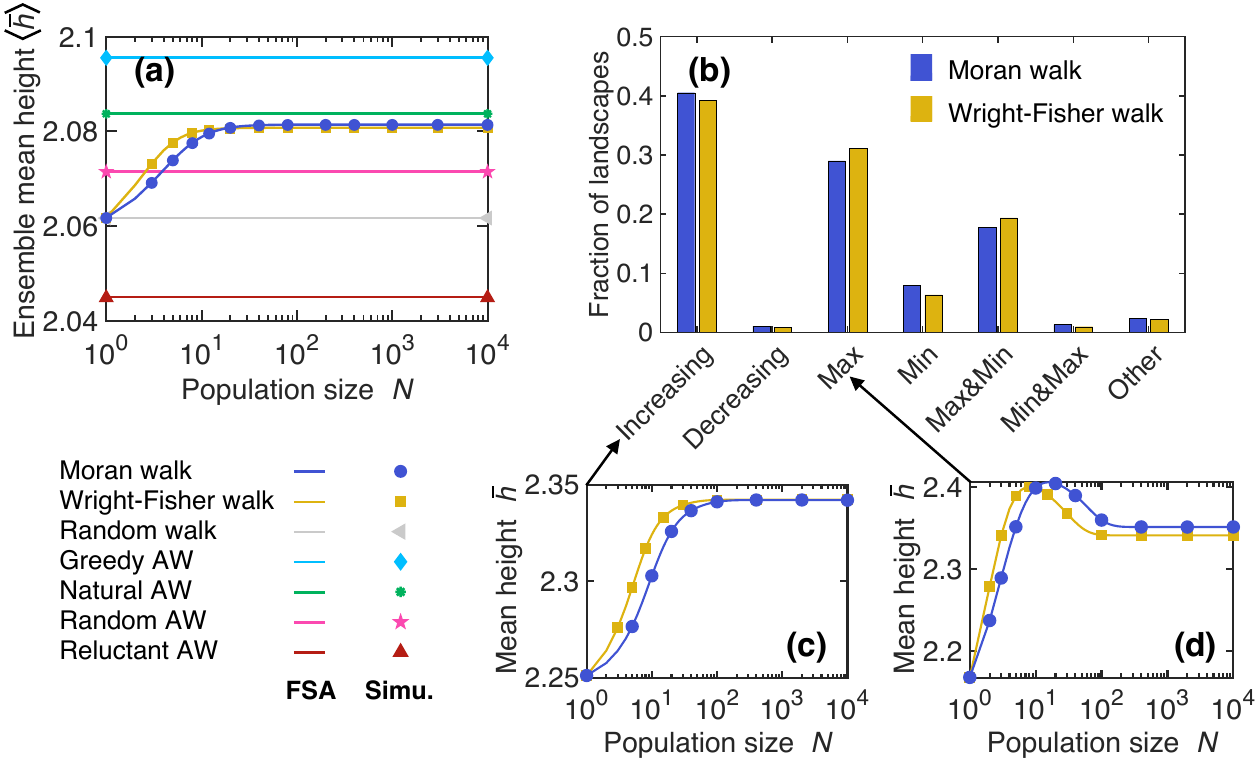}
 \caption{\textbf{Impact of population size on early adaptation in $LK$ landscapes with $L=3$ and $K=1$.} (a) Ensemble mean height $\left\langle \bar{h} \right\rangle$ of the first fitness peak reached when starting from a uniformly chosen initial node versus population size $N$ for various walks. Lines: numerical resolutions of the FSA equations for each landscape; markers: simulation results averaged over $100$ walks per starting node in each landscape. In both cases, the ensemble average is performed over $5.6\times 10^5$ landscapes. (b) Distribution of behaviors displayed by $\bar{h}$ versus $N$ for the Moran and Wright-Fisher walks over $2 \times 10^5$ landscapes. Classes of behaviors of $\bar{h}$ versus $N$ are: monotonically increasing or decreasing, one maximum, one minimum, one maximum followed by a minimum at larger $N$ (``Max\&Min''), vice-versa (``Min\&Max''), and more than two extrema (``Other''). In each landscape, we numerically solve the FSA equations for various $N$. (c-d) $\bar{h}$ versus $N$ is shown in two example landscapes (see Table~\ref{table:landscapes}), landscape A yielding a monotonically increasing behavior (c) and landscape B yielding a maximum (d). Same symbols as in (a); simulation results averaged over $10^5$ (c) and $5\times10^5$ (d)  walks per starting node.}
 \label{fig:average_L=3_K=1}
\end{figure}

\begin{figure}[h!]
 \centering
 \includegraphics[width=0.9\textwidth]{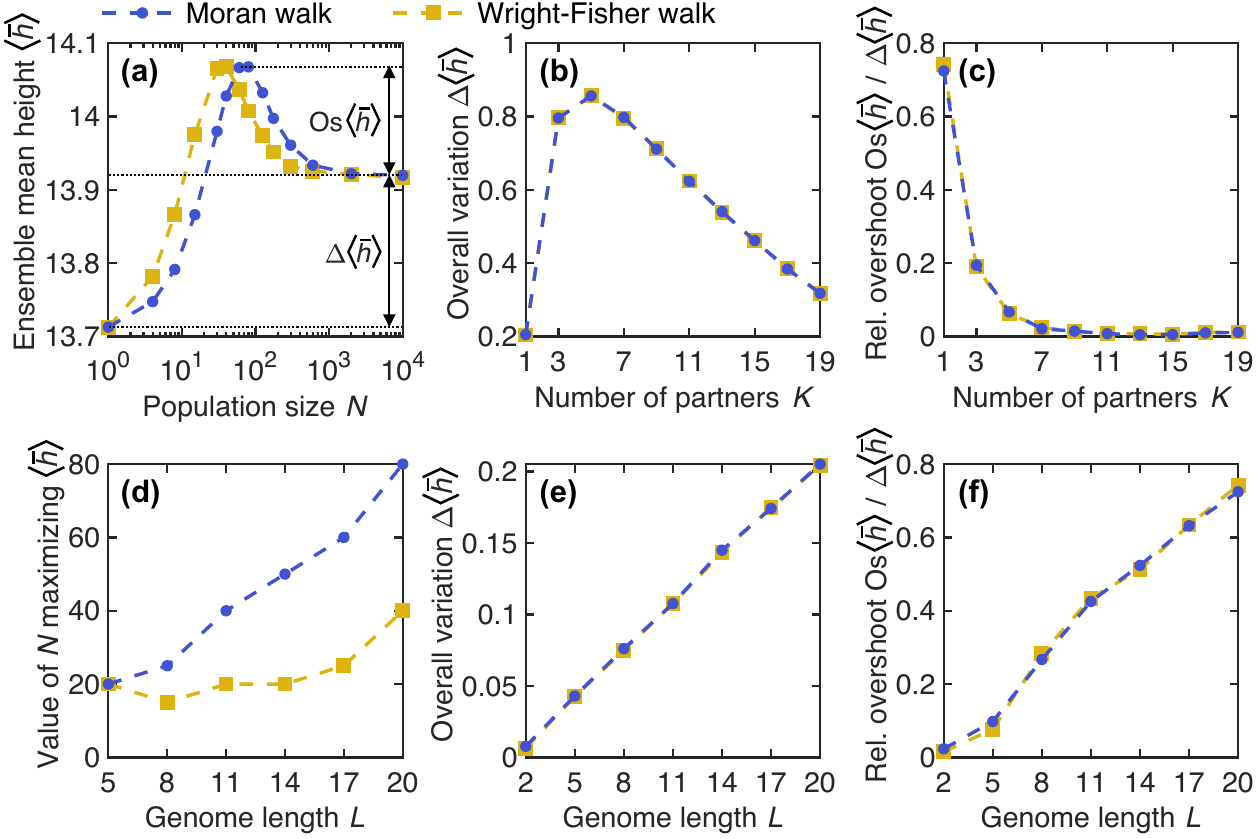}
\caption{\textbf{Impact of $L$ and $K$ on early adaption in $LK$ landscapes.} (a) Ensemble mean height $\left\langle \bar{h} \right\rangle$ of the first fitness peak reached when starting from a uniformly chosen initial node for $LK$ landscapes with $L = 20$ and $K = 1$ versus population size $N$ for the Moran and Wright-Fisher walks. The overall variation $\Delta\!\left<\bar{h}\right>=\left<\bar{h}\right>_\infty - \,\left<\bar{h}\right>(N=1)$ and the overshoot $\textrm{Os}\!\left<\bar{h}\right>=\text{max}\left<\bar{h}\right> - \left<\bar{h}\right>_\infty$ are indicated. (b) Overall variation $\Delta\!\left\langle \bar{h} \right\rangle$ versus the number of partners $K$, for $L=20$. (c) Relative overshoot $\textrm{Os}\!\left\langle \bar{h} \right\rangle/\Delta\!\left\langle \bar{h} \right\rangle$ versus $K$, for $L=20$. (d) Value of the population size $N$ that maximizes $\left\langle \bar{h} \right\rangle$ versus genome length $L$ (i.e.\ number of binary loci), for $K=1$. (e) Overall variation of $\left<\bar{h}\right>$ (as in (b)), versus $L$, for $K=1$. (f) Relative overshoot of $\left< \bar{h} \right>$ (as in (c)), versus $L$, for $K=1$. Markers connected by dashed lines are simulation results from $5\times10^5$ walks ($10^7$ for $L = 2$), each in a different landscape, generated along the way to save memory. The large-$N$ limit $\left<\bar{h}\right>_\infty$ is evaluated for $N=10^4$.}
\label{fig:amplitudes}
\end{figure}

\paragraph{Average over $LK$ landscapes with $L=3$ and $K=1$.} 
\label{subsec:average}
\cref{fig:average_L=3_K=1}(a) shows that the ensemble mean $\left\langle\bar{h}\right\rangle$ of $\bar{h}$ over these landscapes monotonically increases with $N$ both for the Moran and for the Wright-Fisher walk. FSA and stochastic simulation results (see Methods) are in very good agreement. Thus, on average over these landscapes, larger population sizes make early adaptation more efficient. This is intuitive because natural selection becomes more and more important compared to genetic drift as $N$ increases, biasing the walks toward larger fitness increases. \cref{fig:average_L=3_K=1}(a) also shows $\left\langle\bar{h}\right\rangle$ for the various adaptive walks (AWs)~\cite{nowak2015} defined in the Supplementary material, \cref{sec:AW_models}, and for the pure random walk. For $N = 1$, the Moran and Wright-Fisher walks reduce to pure random walks, since all mutations are accepted. For $N\to\infty$, where the Moran and Wright-Fisher walks become AWs, $\left\langle\bar{h}\right\rangle$ is close to the value obtained for the natural AW, where the transition probability from $i$ to $j$ is proportional to $s_{ij}=f_j/f_i-1$ if $s_{ij}>0$ and vanishes if $s_{ij}<0$~\cite{gillespie1983,Gillespie84,Neidhart11}. Indeed, when $N\to\infty$, the Moran (resp. Wright-Fisher) fixation probability in \cref{eq:moran} (resp. \cref{eq:transition_proba_wf}) converges to $s_{ij}/\left(1+s_{ij}\right)$ (resp. $1-\exp(-2s_{ij})$) if $s_{ij}>0$ and to 0 otherwise. If in addition $0<s_{ij}\ll 1$ while $Ns_{ij} \gg 1$, then they converge to $s_{ij}$ and $2s_{ij}$ respectively, and both become equivalent to the natural AW. The slight discrepancy between the asymptotic behavior of the Moran and Wright-Fisher walks and the natural AW comes from the fact that not all $s_{ij}$ satisfy $|s_{ij}|\ll 1$ in these landscapes. Convergence to the large-$N$ limit occurs when $Ns_{ij}\gg 1$ for all relevant $s_{ij}$, meaning that landscapes with near-neutral mutations will feature finite-size effects up to larger $N$. Besides, this convergence occurs for slightly larger $N$ for the Moran walk than for the Wright-Fisher walk (see \cref{fig:average_L=3_K=1}(a)). Indeed, if $s_{ij}>0$, $\ln(1+s_{ij})<2s_{ij}$, so \cref{eq:transition_proba_wf} converges to its large-$N$ limit faster than \cref{eq:moran}, while if $-0.79<s_{ij}<0$, $\ln(1+s_{ij})>2s_{ij}$, so \cref{eq:transition_proba_wf} tends to $0$ faster than \cref{eq:moran} for large $N$ ($s_{ij}<-0.79$ is very rare and yields tiny fixation probabilities). Note that on \cref{fig:average_L=3_K=1}(a), the range of variation of $\left\langle \bar{h} \right\rangle$ with $N$ is small, but this is landscape-dependent (see \cref{fig:amplitudes}).

How does the time needed to reach the first peak depend on $N$ in $LK$ landscapes with $L=3$ and $K=1$? First, the ensemble mean length $\left\langle\bar{\ell}\right\rangle$, defined as the mean number of mutation fixations before the first peak is reached, decreases before saturating as $N$ increases, see \cref{fig:t_and_l_vs_N}(a). Indeed, when $N$ increases, the walk becomes more and more biased toward increasing fitness. Conversely, the ensemble mean time $\left\langle \bar{t}\right\rangle$, defined as the mean number of mutation events (fixing or not) before the first peak is reached, increases with $N$, see \cref{fig:t_and_l_vs_N}(b). Indeed, many mutations are rejected for large $N$. Moreover, at a given $s_{ij}$, fixation probabilities (\cref{eq:moran} and \cref{eq:transition_proba_wf}) decrease as $N$ increases. Note that for $Ns_{ij}\gg 1$ and $s_{ij}\ll 1$, the limit of \cref{eq:moran} is $s_{ij}$ while that of \cref{eq:transition_proba_wf} is $2s_{ij}$, explaining why the large-$N$ limit of $\left\langle \bar{t}\right\rangle$ is about twice larger for the Moran than for the Wright-Fisher walk. Note that, more generally, a factor of 2 differs between the diffusion limits of the fixation probabilities of the Moran and the Wright-Fisher models. It arises from a difference in the variance in offspring number~\cite{ewens1979}. Finally, since mutations occur proportionally to $N$, the actual time needed by the population to reach the first peak is proportional to $\left\langle \bar{t}\right\rangle/N$, which decreases with $N$, see \cref{fig:t_and_l_vs_N}(c).

\paragraph{Diversity among $LK$ landscapes with $L=3$ and $K=1$.}
\label{subsec:diversity}
How much does the population-size dependence of $\bar{h}$ depend on the specific landscape considered? To address this question, we focus on landscapes that have more than one peak (46\% of $L=3$, $K=1$ landscapes), since with a single peak, $\bar{h}$ is always equal to the fitness of that peak. Interestingly, \cref{fig:average_L=3_K=1}(b) shows that $\bar{h}$ does not always monotonically increase with $N$. In fact, this expected case occurs only for about 40\% of the landscapes with more than one peak, see e.g. \cref{fig:average_L=3_K=1}(c), and $\bar{h}$ can exhibit various behaviors versus $N$. Around 30\% of landscapes with more than one peak yield a single maximum of $\bar{h}$ versus $N$, see e.g. \cref{fig:average_L=3_K=1}(d). For these landscapes, there is a specific finite value of $N$ that optimises early adaptation. While some landscapes yield multiple extrema of $\bar{h}$ versus $N$, the absolute amplitude of secondary extrema is generally negligible. Indeed, when $\bar{h}$ versus $N$ displays two or more extrema, the mean ratio of the amplitude of the largest extremum to that of other extrema is larger than 20. Here, the amplitude of the $i$-th extremum starting from $N=1$, observed at $N=N_i$, is computed as the mean of $A_i=|\bar{h}(N_i)-\bar{h}(N_{i-1})|$ and $A_{i+1}$ (where $N_0=1$ and $N_{i+1}\to\infty$ for the last extremum).
\cref{fig:average_L=3_K=1}(b) shows that in 14\% of landscapes with more than one peak, the Moran and the Wright-Fisher walks exhibit different behaviors. However, the scale of these differences is negligible. As illustrated by \cref{fig:average_L=3_K=1}(b), studying the behavior of $\bar{h}$ versus $N$ could be useful to characterize and classify fitness landscapes, and potentially complementary to epistasis measures in~\cite{aita2001,poelwijk2007,szendro2013b,ferretti2016,ferretti2018}. 

The mean length $\bar{\ell}$ and time $\bar{t}$ of the walk also vary across landscapes, but the same overall trends as for the ensemble mean length and time are observed, see \cref{fig:t_and_l_vs_N}(d-i).

\paragraph{Impact of the starting set of genotypes.} So far, we have considered the average $\bar{h}$ of $h$ over all possible initial genotypes, assumed to be equally likely. What is the impact of restricting the set of possible starting points to those with high fitness? This question is relevant to adaptation after small to moderate sudden environmental changes~\cite{orr2005}, where the wild type is no longer optimal, but still has relatively high fitness. To address it, we choose starting points uniformly among the $n$ fittest genotypes. In \cref{fig:h_vs_N_varying_starting_set}, we study the same landscapes as in \cref{fig:average_L=3_K=1}, varying $n$ between 1 and $2^L$. For $n=2^L$, all genotypes can be starting points, as before. The behavior of the ensemble mean $\left\langle\bar{h}\right\rangle$ versus $N$ is similar across the different sets of starting points (\cref{fig:h_vs_N_varying_starting_set}(a)). \cref{fig:h_vs_N_varying_starting_set}(b) further shows that $\bar{h}$ versus $N$ always monotonically increases for the landscape of \cref{fig:average_L=3_K=1}(c). Finally, in \cref{fig:h_vs_N_varying_starting_set}(c), $\bar{h}$ versus $N$ displays an intermediate maximum for the landscape of \cref{fig:average_L=3_K=1}(d), except for $n=1$ and $n=2$. In these cases, the possible starting points are either only the absolute peak of the landscape, or itself and one of its neighbors that has a higher fitness than the small peak, so the latter rarely comes into play. This is also why the values of $\bar{h}$ are substantially larger for $n = 1$ and $n = 2$ than in other cases. Overall, these results suggest that our main conclusions are robust to varying the set of starting genotypes.

\paragraph{Predicting the overall behavior of $\bar{h}$.} Why do different $L=3$, $K=1$ landscapes yield such diverse behaviors of $\bar{h}$ versus $N$? To address this question, let us first focus on the \textit{overall behavior} of $\bar{h}$ versus $N$, i.e.\ on whether $\bar{h}$ is larger for $N\to\infty$ (\textit{overall increasing}) or for $N=1$ (\textit{overall decreasing}). This distinction is robust across the Moran and Wright-Fisher walks, as their overall behavior differs only in $0.4 \%$ of the landscapes with more than one peak.

Let us focus on the landscapes with 2 peaks (99.5\% of the landscapes with more than one peak) for simplicity. $87\%$ of them yield an overall increasing behavior of $\bar{h}$ versus $N$, as e.g. those featured in \cref{fig:average_L=3_K=1}(c) and (d). Intuitively, the higher a peak, the more attractive it becomes for large $N$ given the larger beneficial mutations leading to it, and an overall increasing behavior is thus expected. However, the opposite might happen if more paths with only beneficial mutations lead to the low peak than to the high peak -- the low peak is then said to be \textit{more accessible} than the high peak. Indeed, when $N \rightarrow \infty$, only beneficial mutations can fix. Therefore, we compare the accessibility of the high peak and of the low peak. 

\begin{figure}[h!]
 \centering
 \includegraphics[width=\textwidth]{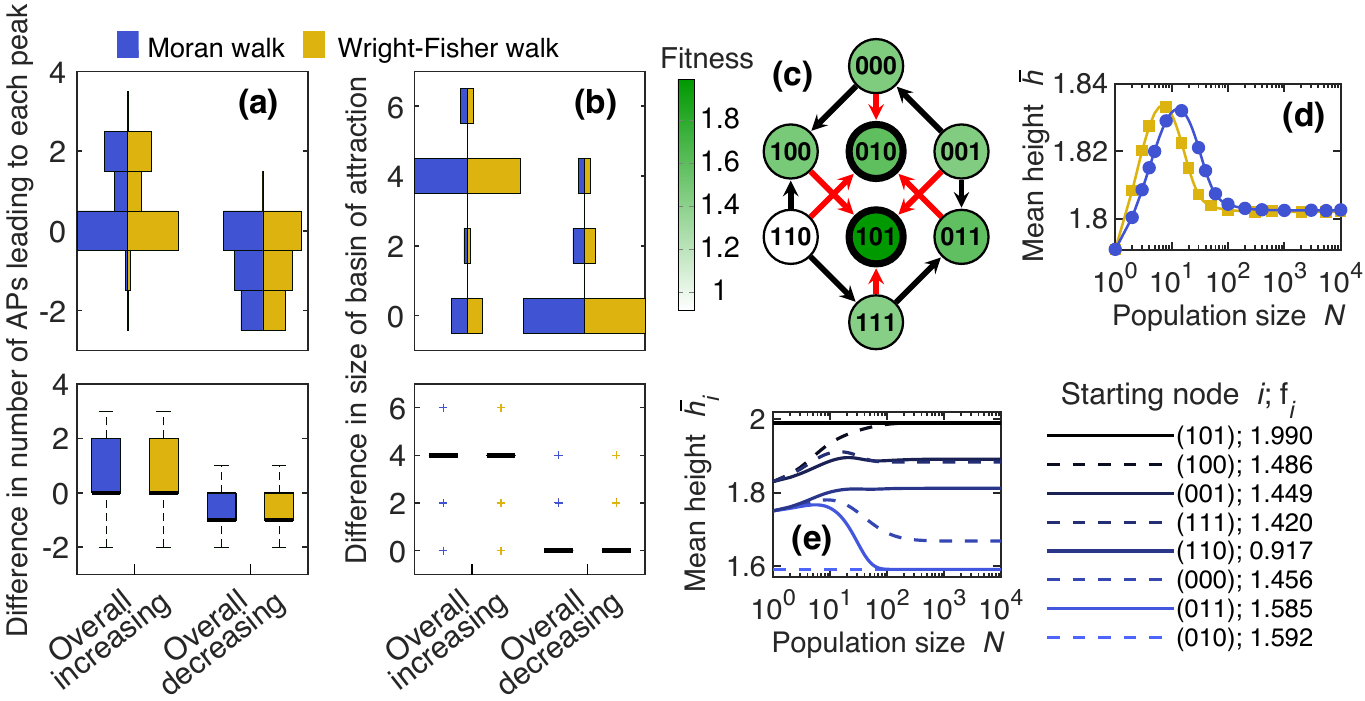}
 \caption{\textbf{Accessibility of peaks and overall population-size dependence of early adaptation in $LK$ landscapes with $L=3$ and $K=1$.} All $9.2\times 10^4$ 2-peak landscapes from an ensemble of $2\times10^5$ landscapes were sorted according to whether $\bar{h}$ versus $N$ displays an overall increasing or decreasing behavior (for the Moran and Wright-Fisher walks). (a-b) Distributions of two measures of differential accessibility of the high and low peaks (see main text) for these two classes of landscapes. Top panels: histograms (displayed vertically); bottom panels: associated box plots (bold black line: median; colored boxes: 25th and 75th percentiles; dashed lines: minimum and maximum values that are not outliers; crosses: outliers). (c) Example landscape where both differential accessibility measures in (a-b) are 0. Bold circled nodes are peaks; arrows point towards fitter neighbors; red arrows point toward fittest neighbors. (d) $\bar h$ versus $N$ for the Moran and Wright-Fisher walks in the landscape in (c). (e) Mean height $\bar{h_i}$ starting from each node $i$ versus $N$ for the Moran walk in the landscape in (c). Lines (d-e): numerical resolutions of FSA equations; markers (d): simulation results averaged over $10^5$ walks per starting node.}
\label{fig:accessibility_measures}
\end{figure}

In \cref{fig:accessibility_measures}(a-b), we show the distributions of two measures reflecting this differential accessibility in the 2-peak landscapes with either overall increasing or overall decreasing dependence of $\bar{h}$ on $N$. The first measure (\cref{fig:accessibility_measures}(a)) is the number of accessible paths (APs)~\cite{Franke11} leading to the high peak minus the number of those leading to the low peak, where APs are paths comprising only beneficial mutations (note that APs included in other APs are not counted). The second measure (\cref{fig:accessibility_measures}(b)) is the size of the basin of attraction of the high peak minus that of the low peak, where the basin of attraction is the set of nodes from which a greedy AW, where the fittest neighbor is chosen at each step, leads to the peak considered~\cite{visser2009,franke2012}. \cref{fig:accessibility_measures} shows that landscapes displaying overall increasing behaviors tend to have a high peak more accessible than the low peak, and vice-versa for landscapes displaying overall decreasing behaviors. Quantitatively, 99\% of the landscapes where both measures are positive or 0, but not both 0 (representing 75.2\% of 2-peak landscapes), yield an overall increasing behavior. Moreover, 91\% of the landscapes where both measures are negative or 0, but not both 0 (representing 5.7\% of 2-peak landscapes), yield an overall decreasing behavior. Hence, differential accessibility is a good predictor of the overall behavior of $\bar{h}$ versus~$N$. Note that combining both measures is substantially more precise than using either of them separately (for instance, landscapes where the AP-based measure is strictly negative yield only 73\% of overall decreasing behaviors).

However, for 15.2\% of all 2-peak landscapes, both differential accessibility measures are 0, and thus do not predict the overall behavior. In practice, 70\% of these tricky landscapes yield an overall increasing behavior. One of those is shown in \cref{fig:accessibility_measures}(c)  and in a complementary representation in \cref{fig:landscape_fig2_magellan}. It yields the $\bar{h}$ versus $N$ curve in \cref{fig:accessibility_measures}(d). Note that another tricky, but rarer case, corresponds to landscapes where accessibility measures have strictly opposite signs (3.9\% of 2-peak landscapes). 

\paragraph{Finite-size effects on $\bar{h}$ and $\left<\bar{h}\right>$.} Predicting the intermediate-$N$ behavior, e.g. the maximum in \cref{fig:accessibility_measures}(d), is more difficult than predicting the overall behavior, and our accessibility measures do not suffice for this, nor do various ruggedness and epistasis measures from~\cite{aita2001,poelwijk2007,szendro2013b,ferretti2016,ferretti2018}. To understand this, let us consider the landscape in \cref{fig:accessibility_measures}(c). The mean heights $\bar{h}_i$ starting from each node $i$ are displayed in \cref{fig:accessibility_measures}(e), showing that diverse behaviors combine to give that of $\bar{h}$. Starting from node $(011)$, the only accessible path is to the low peak $(010)$, so $\bar{h}_i$ decreases when $N$ increases, but the quite small differences of fitnesses between $(011)$ and its neighbors mean that relatively large values of $N$ are required before this matters. Indeed, the convergence of fixation probabilities to their large-$N$ limits occurs when $N |s_{ij}|\gg 1$ (see \cref{eq:moran} and \cref{eq:transition_proba_wf}). Conversely, starting from $(100)$, the only accessible path is to the high peak $(101)$, so $\bar{h}_i$ increases with $N$, starting at smaller values of $N$ due to the larger fitness differences involved, e.g. between $(100)$ and $(110)$. Such subtle behaviors, which depend on exact fitness values in addition to peak accessibility, yield the maximum in \cref{fig:accessibility_measures}(d).

For landscape B (\cref{fig:average_L=3_K=1}(d)), which also yields a maximum of $\bar{h}$ at an intermediate value of $N$, \cref{fig:sd_h} shows that the standard deviation $\sigma_h$ of the height $h$ reached from a uniformly chosen starting node features a minimum at a similar $N$, while the average $\overline{\sigma_{h_i}}$ over starting nodes $i$ of the standard deviation of the height $h_i$ starting from node $i$ monotonically decreases when $N$ increases. This corroborates the importance of the diversity of behaviors with starting nodes $i$ in the finite-size effects observed. Moreover, the minimum in the standard deviation $\sigma_h$ starting from any node means that early adaptation is more predictable for intermediate values of $N$. Note that $\sigma_h$ and $\overline{\sigma_{h_i}}$ both decrease with $N$ for landscape A, where $\bar{h}$ increases with $N$ (\cref{fig:average_L=3_K=1}(c)), see \cref{fig:sd_h}. 

\paragraph{Magnitude of the overall variation of $\bar{h}$.} We showed above that for two-peak fitness landscapes with $L=3$ and $K=1$, the differential accessibility of the peaks allows to predict the overall behavior of $\bar{h}$, i.e.\ the sign of $\Delta\bar{h}=\bar{h}_\infty-\bar{h}(N=1)$, where $\bar{h}_\infty$ denotes the large-$N$ limit of $\bar{h}$. What determines the magnitude of $\Delta\bar{h}$? \cref{fig:sd_fitness_of_the_peaks} shows that it strongly correlates with the standard deviation $\sigma_{f_M}$ of the peak fitness values. This makes sense, as the range of $\bar h$ in a landscape is bounded by the fitness of the lowest peak and that of the highest peak. 

\paragraph{Impact of $L$ and $K$.} So far we focused on small $LK$ landscapes with $L=3$ and $K=1$, which generally have one or two peaks. However, real fitness landscapes generally involve much larger genome lengths $L$ (number of binary units, representing genes, nucleotides, or amino acids) and may involve larger numbers of epistatic partners $K$ and be more rugged. How do these two parameters impact $\left\langle \bar{h}\right\rangle$? First, $\left\langle \bar{h}\right\rangle$ increases linearly with $L$ at $K=1$, because all fitness values increase linearly with $L$ in $LK$ landscapes (\cref{eq:$LK$_model}), see \cref{fig:vsL_K}(a). For adaptive walks in $LK$ landscapes, such a linear behavior was analytically predicted with block neighborhoods and holds more generally when $L\gg K$~\cite{nowak2015}. We also find that for each value of $N$, $\left\langle \bar{h}\right\rangle$ features a maximum for an intermediate $K$ at $L=20$, see \cref{fig:vsL_K}(b).  A similar observation was made on adaptive walks in~\cite{nowak2015}. We find that the value of $K$ that maximizes $\left\langle \bar{h}\right\rangle$ depends on $N$, illustrating the importance of epistasis for finite-size effects.

While $\left\langle \bar{h}\right\rangle$ monotonically increases with $N$ for $L=3$ and $K=1$ (\cref{fig:average_L=3_K=1}(a)), a pronounced maximum appears at finite $N$ for larger $L$, see \cref{fig:amplitudes}(a). To quantify how $\left\langle \bar{h}\right\rangle$ changes with $N$, we consider the overall variation $\Delta\!\left<\bar{h}\right> =\left<\bar{h}\right>_\infty - \,\left<\bar{h}\right>(N=1)$ of $\left\langle \bar{h}\right\rangle$ between $N=1$ and the large-$N$ limit, as well as the overshoot of the large-$N$ limit, $\textrm{Os}\!\left<\bar{h}\right>=\text{max}\left<\bar{h}\right> - \left<\bar{h}\right>_\infty$, see \cref{fig:amplitudes}(a). Their dependence on $K$ and $L$ is studied in \cref{fig:amplitudes}. First, \cref{fig:amplitudes}(b) shows that for $L=20$, $\Delta\!\left<\bar{h}\right>$ is maximal for $K=5$, while \cref{fig:amplitudes}(c) shows that the relative overshoot $\textrm{Os}\!\left<\bar{h}\right>/\Delta\!\left<\bar{h}\right>$ is maximal for $K=1$, and rapidly decreases for higher $K$. This is interesting, as $K=1$ corresponds to pairwise interactions, highly relevant in protein sequences~\cite{Weigt09,Marks11,Morcos11}. Next, we varied $L$ systematically for $K=1$ (see \cref{fig:h_vs_N_multiple_L} for examples). A maximum of $\left\langle \bar{h}\right\rangle$ at finite $N$ exists for $L=4$ and above. The associated value of $N$ increases with $L$ (see \cref{fig:amplitudes}(d)), but it remains modest for the values of $L$ considered here. A key reason why finite-size effects matter for larger $N$ when $L$ increases is that more mutations are then effectively neutral, i.e.\ satisfy $N|s_{ij}|\ll 1$. This abundance of effectively neutral mutations is relevant in natural situations~\cite{robert2018mutation}. Furthermore, \cref{fig:amplitudes}(e) shows that $\Delta\!\left<\bar{h}\right>$ increases with $L$, and \cref{fig:amplitudes}(f) shows that $\textrm{Os}\!\left<\bar{h}\right>/\Delta\!\left<\bar{h}\right>$ also increases with $L$, exceeding 0.7 for $L=20$. Thus, finite-size effects on early adaptation become more and more important as $L$ is increased. This hints at important possible effects in real fitness landscapes, since genomes have many units (genes or nucleotides).   
As shown in \cref{fig:peak_density_L}, the density of peaks in the landscapes considered here decreases when $L$ increases at $K=1$, consistently with analytical results for large $L$ and $K$~\cite{hwang2018}. Therefore, maxima of $\bar{h}$ versus $N$ are associated to rugged landscapes with sparse peaks.

Beyond these ensemble mean behaviors, we analyze how $L$ and $K$ impact the diversity of behaviors of $\bar{h}$ with $N$ in \cref{fig:class_proportions}. We find that the proportion of landscapes (with more than one peak) yielding a monotonically increasing $\bar{h}$ with $N$ decreases steeply as $L$ increases when $K=1$, while the proportion of those yielding a maximum at intermediate $N$ increases. An opposite, but less steep, trend is observed as $K$ increases at $L=6$. Thus, most landscapes yield a maximum of $\bar{h}$ at finite $N$ when $L$ is large enough and $K=1$ -- the maximum of $\left\langle \bar{h} \right\rangle$ does not just arise from averaging over many landscapes. Besides, about 10\% of landscapes with more than one peak yield an overall decreasing behavior of $\bar{h}$ with $N$ (i.e., $\bar{h}_\infty<\bar{h}(N=1)$) when $K=1$ for all $L>2$ considered, while this proportion decreases if $K$ increases at $L=6$.

Finally, the impact of $L$ and $K$ on $\left\langle\bar{\ell}\right\rangle$ is shown in~\cref{fig:vsL_K}(c-d): $\left\langle\bar{\ell}\right\rangle$ increases with $L$ for $K=1$, more strongly if $N$ is small, and $\left\langle\bar{\ell}\right\rangle$ decreases as $K$ increases at $L=20$. Indeed, a larger $K$ at constant $L$ entails more numerous peaks and a larger $L$ at constant $K$ yields smaller peak density (the number of peaks increases less fast with $L$ than the number of nodes). In addition, smaller $N$ means more wandering in the landscapes and larger $\left\langle\bar{\ell}\right\rangle$. Note that for adaptive walks in $LK$ landscapes, a linear behavior of $\left\langle\bar{\ell}\right\rangle$ versus $L$ was analytically predicted with block neighborhoods and holds more generally for $L\gg K$~\cite{nowak2015}. 

\paragraph{Impact of the neighborhood scheme.} So far, we considered the random neighborhood scheme~\cite{nowak2015} where epistatic partners are chosen uniformly at random. We find qualitatively similar behaviors in two other neighborhood schemes, see \cref{fig:neighborhoods}.

\subsection{Extension to various model and experimental fitness landscapes}
\label{subsec:overview}

While the $LK$ model is convenient as it allows to explicitly tune epistasis and ruggedness, many other models exist, and natural fitness landscapes have been measured~\cite{Fragata19}. How general are our findings on the population-size dependence of early adaptation across fitness landscapes?

\paragraph{Model fitness landscapes.} We first consider different landscape models (see Supplementary material, \cref{sec:landscape_models}). In all of them, $\left\langle \bar{h} \right\rangle$ is overall increasing between $N=1$ and the large-$N$ limit, and either monotonically increases or features a maximum at intermediate $N$. This is consistent with our findings for $LK$ landscapes, demonstrating their robustness. Specifically, we find maxima of $\left\langle \bar{h} \right\rangle$ for the $LKp$ model, which includes neutral mutations~\cite{barnett1998}, for the $LK$ model with more than two states per site~\cite{zagorski2016}, and for the Ising model~\cite{diu1989,ferretti2016}, see \cref{fig:overview_models}(a-c). Conversely, in models with stronger ruggedness (House of Cards landscapes, Rough Mount Fuji landscapes~\cite{aita2000,szendro2013b} with strong epistatic contributions and  Eggbox landscapes~\cite{ferretti2016}), we observe a monotonically increasing $\left\langle \bar{h} \right\rangle$, see \cref{fig:overview_models}(d-f). \cref{fig:peak_density_models} shows that the density of peaks is generally smaller than 0.1 in the first three landscape ensembles and larger in the last three. This is consistent with our results for $LK$ landscapes with $K=1$ and different $L$ (see above and \cref{fig:peak_density_L}), confirming that maxima of $\bar{h}$ versus $N$ are associated to rugged landscapes with sparse peaks. Note that tuning the parameters of the model landscape ensembles considered here can yield various peak densities and behaviors, which we did not explore exhaustively.

\begin{figure}[h!]
 \centering
 \includegraphics[width=\textwidth]{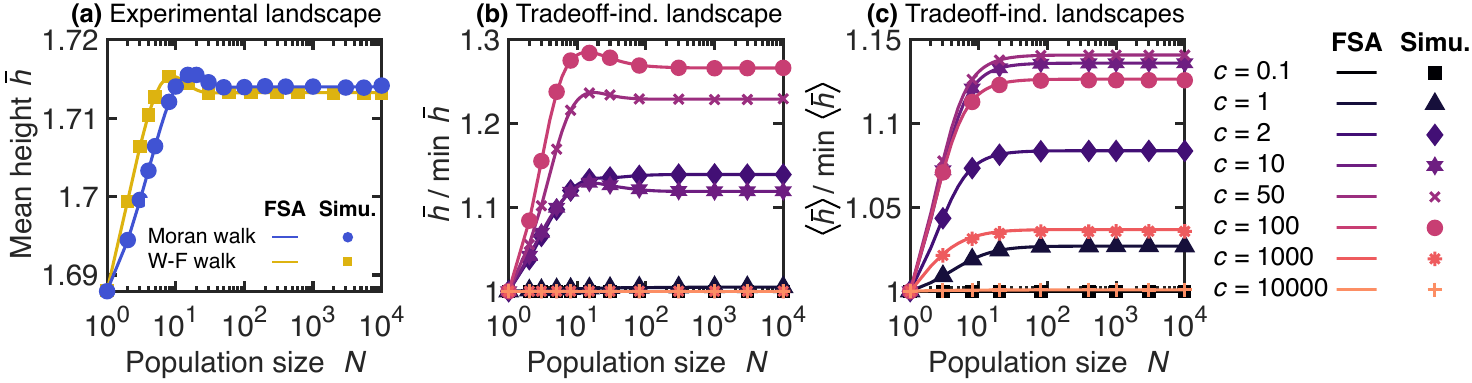}
 \caption{\textbf{Impact of population size on early adaptation in an experimental landscape and in tradeoff-induced landscapes.} (a) Mean height $\bar{h}$ versus population size $N$, for the Moran and Wright-Fisher walks, in the experimental landscape from Table S2 in~\cite{lozovsky2009}. In this study, the fitness landscape of \textit{Escherichia coli} carrying dihydrofolate reductase from  the malaria parasite \textit{Plasmodium falciparum} was measured experimentally with or without the drug pyrimethamine. The focus was on 4 (wild type or mutant) amino acids in dihydrofolate reductase that are important for pyrimethamine resistance, yielding a binary landscape with $L = 4$. The landscape studied here, without pyrimethamine, possesses two peaks. (b) Mean height $\bar{h}$ of the Moran walk, normalized by its minimal value, versus $N$, in a tradeoff-induced landscape~\cite{das2020} with $L=6$ (see \cref{table:r_m_TIL}), for various (dimensionless) antibiotic concentrations $c$. (c) Ensemble mean height $\left\langle\bar{h}\right\rangle$ of the Moran walk, normalized by its minimal value, versus $N$, in an ensemble of tradeoff-induced landscapes~\cite{das2020} with $L=6$ (see Supplementary material, \cref{sec:landscape_models}), for various $c$. Lines: numerical resolutions of FSA equations; markers: simulation data averaged over $10^5$ (a) and $10^4$ (b) walks per starting node. In (c), one walk is simulated per starting node in each landscape, and the ensemble average is over $1.5\times10^6$ (resp. $5.6\times10^5$) landscapes for simulations (resp. FSA) ($10^5$ for $c=0.1$, $10^3$ and $10^4$).}
 \label{fig:exp_TIL}
\end{figure}

\paragraph{Experimental and experimentally-motivated landscapes.} We study $\bar{h}$ versus $N$ in 8 experimental rugged landscapes, see \cref{fig:exp_TIL}(a) and \cref{fig:experimental_landscapes}. In all cases, we observe an overall increasing behavior, most of them generally increasing, and two with a notable maximum at an intermediate size $N$, see \cref{fig:exp_TIL}(a) and \cref{fig:experimental_landscapes}(a). This is consistent with our results for model fitness landscapes, and shows their generality.

Tradeoff-induced landscapes were introduced to model the impact of antibiotic resistance mutations in bacteria, in particular their tendency to increase fitness at high antibiotic concentration but decrease fitness without antibiotic~\cite{das2020,dasPreprint}, see Supplementary material, \cref{sec:landscape_models}. These landscapes tend to be smooth at low and high antibiotic concentrations, but more rugged at intermediate ones, due to the tradeoff~\cite{das2020}. In a specific tradeoff-induced landscape, we find that $\bar{h}$ versus $N$ is flat for the smallest concentrations considered (the landscape has only one peak), becomes monotonically increasing for larger ones, and exhibits a maximum for even larger ones, before becoming flat again at very large concentrations, see \cref{fig:exp_TIL}(b). In all cases, $\bar{h}$ versus $N$ is overall increasing or flat. Besides, the ensemble average over a class of tradeoff-induced landscapes (see Supplementary material, \cref{sec:landscape_models}) yields monotonically increasing behaviours of $\left\langle\bar{h}\right\rangle$ versus $N$ for most concentrations, except the very small or large ones where it is flat, see \cref{fig:exp_TIL}(c). The overall variation $\Delta\!\left\langle\bar{h}\right\rangle$ is largest (compared to the minimal value of $\left\langle\bar{h}\right\rangle$) for intermediate concentrations. These findings are consistent with our results for model and experimental fitness landscapes, further showing their generality.

\section{Discussion}
\label{sec:discussion}

We studied early adaptation of finite populations in rugged fitness landscapes in the weak mutation regime, starting from a random genotype. We found that the mean fitness $\bar{h}$ of the first encountered peak depends on population size $N$ in a non-trivial way, in contrast to the steady-state fitness which monotonically increases with $N$. We showed that the accessibility of different peaks plays a crucial part in whether $\bar{h}$ is larger in the large-$N$ limit or for $N=1$ in simple two-peaked landscapes. A key reason why $\bar{h}$ may not monotonically increase with $N$ is that as $N$ increases, Moran and Wright-Fisher walks lose possible paths as the fixation probability of deleterious mutations vanishes, while also becoming more biased toward larger fitness increases. These two conflicting effects of increasing $N$ yield a tradeoff. Accordingly, we observed that $\bar{h}$ versus $N$ (and even the ensemble mean $\left\langle\bar{h}\right\rangle$) often features a maximum for intermediate $N$, especially in rugged fitness landscapes with small peak densities, where most nodes are relatively far from peaks. In these cases, early adaptation is more efficient, in the sense that higher peaks are found, for intermediate $N$ than in the large-$N$ limit. Studying the behavior of $\bar{h}$ versus $N$ could potentially be useful to characterize and classify landscapes.

Our results hold both for the Moran model, and for the Wright-Fisher model in the diffusion limit. Furthermore, they extend to various model rugged landscapes and to many experimental and experimentally-motivated ones, including several experimental fitness landscapes involved in the evolution of antimicrobial resistance. This shows the robustness of our conclusions and their relevance to biologically relevant situations. 

The time it takes to cross a fitness valley~\cite{Weissman09,Weissman10} and the entropy of trajectories on fitness landscapes~\cite{szendro2013a} depend non-monotonically on $N$. However, both results arise from the possibility of observing double mutants in a wild-type population when $N$ increases at fixed mutation rate $\mu$. Small populations can also yield faster adaptation that larger ones~\cite{Rozen08,Jain11}, but this occurs at the onset of clonal interference. By contrast, we remained in the weak mutation regime, highlighting that even then, population size has non-trivial effects on adaptation. Our focus on weak mutation without strong selection (see also~\cite{Berg04,sella2005,mccandlish2011,mccandlish2014,mccandlish2018}) complements the study of strong selection with frequent mutation~\cite{Laessig17}, going beyond the strong selection weak mutation regime.

The overshoot we find of the large-$N$ limit of $\bar{h}$ is often small. In addition, it occurs for modest values of $N$, meaning that adaptation becomes most efficient for sizes that are quite small compared to the total size of many microbial populations. However, the relative amplitude of the overshoot, and the $N$ at which it occurs, both increase with genome size $L$ in $LK$ landscapes. The large-$L$ case is biologically relevant since genomes have many units (genes or nucleotides). Furthermore, in $LK$ landscapes, the relative overshoot is largest for $K=1$, i.e.\ pairwise epistasis, a case that describes well protein sequence data~\cite{Weigt09,Morcos11,Marks11}. More generally, finite-size effects in early adaptation are expected for population sizes $N$ such that $N|s|$ is small for a sufficient fraction of mutations in the landscape, where $s$ denotes the relative fitness effect of a mutation. Thus, finite-size effects should matter for larger population sizes if neutral and effectively neutral~\cite{ewens1979} mutations are abundant. This is a biologically relevant situation~\cite{robert2018mutation}. 

Besides, spatial structure and population bottlenecks yield smaller effective population sizes, for which our findings are relevant. Studying the effect of spatial structure on early adaptation in rugged fitness landscapes is an interesting topic for future work. Indeed, complex spatial structures with asymmetric updates or migrations impact the probabilities of fixation of mutations~\cite{Lieberman05,Houchmandzadeh11,Yagoobi21,Marrec21,ChakrabortyPreprint}, which should affect early adaptation. Beyond the weak mutation regime, fitness valley crossing by tunneling can aid adaptation~\cite{Weissman09,Weissman10}, which may especially impact subdivided populations, as first discussed in Wright's shifting balance theory~\cite{Wright31,wright1932} and shown in a minimal model~\cite{Bitbol14}. Another interesting direction regards the effect of environment-induced modifications of fitness landscapes on adaptation~\cite{Hall19,Fragata19}.

\section*{Acknowledgments}
 The authors thank Alia Abbara for helpful discussions about Markov chains and Claudia Bank for providing useful data from~\cite{bank2016}. This project has received funding from the European Research Council (ERC) under the European Union’s Horizon 2020 research and innovation programme (grant agreement No.~851173, to A.-F.~B.).


\clearpage
\newpage

\begin{center}
 {\LARGE \bf Supplementary material}   
\end{center}


\renewcommand{\thesection}{S\arabic{section}}
\renewcommand{\theHsection}{S\arabic{section}}
\setcounter{section}{0}
\renewcommand{\thefigure}{S\arabic{figure}}
\setcounter{figure}{0}
\renewcommand{\thetable}{S\arabic{table}}
\setcounter{table}{0} 

\section{Mean fitness evolution and steady state}
\label{sec:steadyfit}

\subsection{Master equation and mean fitness evolution}
\label{subsec:master_equation}

As the Moran and Wright-Fisher walks are Markov chains, one can write a master equation on the probability $P_i$ that the population is in state $i$ (representing its genotype). Let us denote time, expressed in number of mutation events, by a discrete variable $\tau$. The master equation reads:
\begin{equation}
    P_i(\tau+1)-P_i(\tau) = \sum_{j \in G_i} P_j(\tau) \frac{1}{L}P_{ji} - P_i(\tau)\sum_{j \in G_i} \frac{1}{L}P_{ij}\,,
\label{eq:master_eq}
\end{equation}
where $G_i$ is the set of neighbors of $i$ (i.e.\ the $L$ genotypes that differ from $i$ by only one mutation) and the $P_{ij}/L$ are the transition probabilities. Indeed, $1/L$ is the probability that the mutation yields the neighbor $j$ of $i$, while $P_{ij}$ is the fixation probability of this mutation, given by \cref{eq:moran} for the Moran walk or \cref{eq:transition_proba_wf} for the Wright-Fisher walk.

\begin{figure}[h!]
 \centering
 \includegraphics[width=\textwidth]{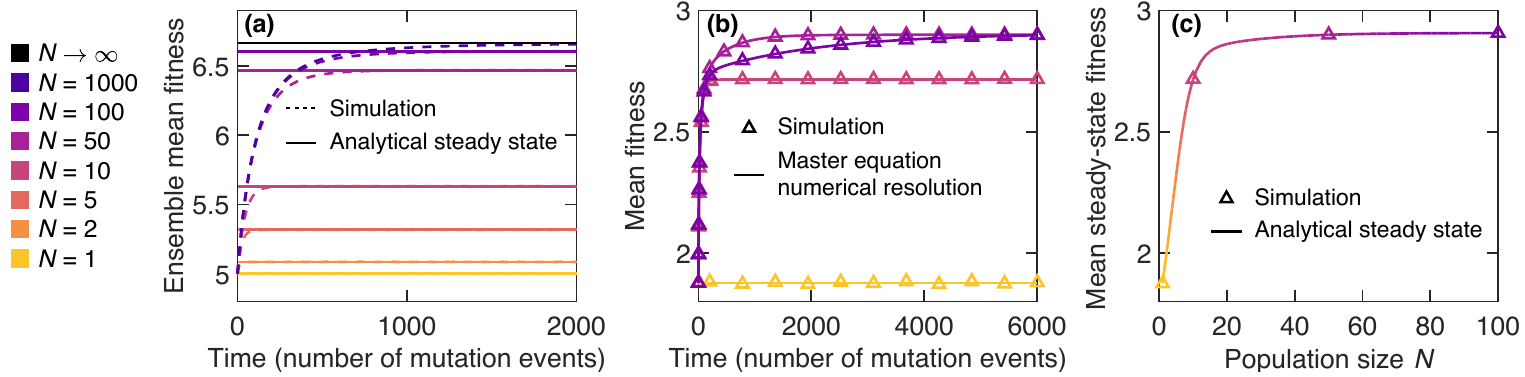}
 \caption{\textbf{Impact of population size on the convergence of the mean fitness to a steady-state value.} (a) Ensemble mean fitness versus time (in number of mutation events) for Moran walks in $LK$ fitness landscapes with $L=10$ and $K=0$, starting from uniformly chosen nodes. Simulation results (dashed lines) and analytical predictions for the steady state (\cref{eq:stationary_mean_fit_2}; solid lines) are shown for different population sizes $N$. In simulations, the ensemble mean is taken by averaging over $1.6\times10^6$ walks, each walk taking place in a different landscape, generated along the way to save memory. The analytical steady-state values are averaged over $2.8\times10^5$ landscapes. (b) Mean fitness versus time for Moran walks in a specific $LK$ fitness landscape with $L=4$ and $K=2$, starting from uniformly chosen nodes. Simulation results (markers), as well as computations of the mean fitness by \cref{eq:dyn_mean_fit}, using a numerical resolution of the Master equation \cref{eq:master_eq} (solid lines), are shown for different $N$. Results are averaged over $10^5$ walks, each of them starting from a uniformly chosen node. (c) Mean steady-state fitness versus population size $N$ in the same fitness landscape as in panel (b). Simulation results (markers) and analytical predictions (\cref{eq:stationary_mean_fit_2}; solid lines) are shown. Simulation results are taken at time $20,000$ to make sure steady-state is reached. }
 \label{fig:fitness_vs_time}
\end{figure}

Solving numerically \cref{eq:master_eq} allows us to compute the mean fitness $F(\tau)$, see \cref{fig:fitness_vs_time}:
\begin{equation}
 F(\tau) = \sum_{i \in G} f_i P_i(\tau)\,,
\label{eq:dyn_mean_fit}
\end{equation}
where $G$ is the set of all nodes and $f_i$ is the fitness of node $i$.

\subsection{Mean steady-state fitness increases with population size}
\label{subsec:steadyfit}

At steady state, the mean fitness in a given fitness landscape, corresponding to the large-$\tau$ limit of \cref{eq:dyn_mean_fit}, is given by
\begin{equation}
 F =\lim_{\tau\to\infty}F(\tau)= \sum_{i \in G} f_i \pi_i\,,
\label{eq:stationary_mean_fit}
\end{equation}
where $G$ is the set of all nodes and $f_i$ is the fitness of $i$, as above, while $\pi_i$ is the stationary probability that the population is in state $i$. Indeed, because the Markov chain corresponding to the Moran or Wright-Fisher walk is irreducible, aperiodic and positive recurrent, it possesses a unique stationary distribution $\pi_i$ towards which it converges for any initial condition~\cite{norris,aldous2002}, and we can write $\lim_{\tau\to\infty}P_i(\tau)=\pi_i$. 

For the Moran process, $\pi_i = f_i^{N-1} / \sum_{k \in G} f_k^{N-1}$~\cite{sella2005}, and thus \cref{eq:stationary_mean_fit} gives:
\begin{equation}
    F = \frac{\sum_{i \in G} f_i^N}{\sum_{i \in G} f_i^{N-1}} = \frac{g_N}{g_{N-1}}\,,
    \label{eq:stationary_mean_fit_2}
\end{equation}
where we introduced the sequence $(g_N)_{N\in\mathbb{N}^*}$ defined by
\begin{equation}
g_N= \sum_{i \in G} f_i^N\,,
\end{equation}
for all positive integer $N$.

To determine how $F$ varies with population size $N$, let us index $F$ by $N$ and study the sign of
\begin{equation}
F_{N+1}-F_{N}=\frac{g_{N+1}}{g_N}-\frac{g_N}{g_{N-1}}=\frac{g_{N+1}g_{N-1}-g_N^2}{g_Ng_{N-1}}\,.
\label{diffavf}
\end{equation}
Because fitness values are positive, $g_N$ is also positive for all positive $N$, and the sign of $F_{N+1}-F_{N}$ is the same as that of its numerator $g_{N+1}g_{N-1}-g_N^2$. Let us thus focus on this quantity:
\begin{equation}
    g_{N+1}g_{N-1}-g_N^2=\sum_{i \in G} f_i^{N+1} \sum_{j \in G} f_j^{N-1}-\sum_{i \in G} f_i^{N} \sum_{j \in G} f_j^{N}=\sum_{(i,j)\in G^2}f_i^Nf_j^N\frac{1}{f_j}\left(f_i-f_j\right)\,.
\label{eq:eq4}
\end{equation}
If $i = j$, $f_i - f_j = 0$. For the remaining terms with $i\neq j$, we can separate the case where $i < j$ and the one where $j > i$, yielding
\begin{align}
    g_{N+1}g_{N-1}-g_N^2 &= \sum_{i < j} f_i^Nf_j^N\frac{1}{f_j}\left(f_i-f_j\right) + \sum_{i > j} f_i^Nf_j^N\frac{1}{f_j}\left(f_i-f_j\right)\nonumber\\
    &=\sum_{i < j} f_i^Nf_j^N\frac{1}{f_j}\left(f_i-f_j\right) + \sum_{j > i} f_i^Nf_j^N\frac{1}{f_i}\left(f_j-f_i\right)\nonumber\\
    &=\sum_{i < j} f_i^Nf_j^N\left(f_i-f_j\right) \left(\frac{1}{f_j}-\frac{1}{f_i}\right)\nonumber\\
    &=\sum_{i < j} f_i^{N-1}f_j^{N-1}\left(f_i-f_j\right)^2\geq 0\,.
\label{eq:derivative_num}
\end{align}

Combining \cref{diffavf} and \cref{eq:derivative_num}, we have shown that for all positive integers $N$, $F_{N+1}-F_{N}\geq 0$, which entails that the mean steady-state fitness $F$ increases with $N$.

\section{Mean length $\bar{\ell}$ and time $\bar{t}$ of a walk}
\label{sec:length_time}

We define the time $t$ of a walk as the total number of mutations (that fix or not) that occur before the first fitness peak is reached. Similarly, the length $\ell$ of a walk is defined as the number of successful fixations that occur before the first fitness peak is reached.

In a given landscape, the mean time $\bar{t}$ (resp. length $\bar{\ell}$) of a walk starting from a uniformly chosen node can be expressed as the average over all starting nodes of the mean time $\bar{t_{i}}$ (resp. length $\bar{\ell_{i}}$) to reach the set $M$ of all peaks starting from node $i$:
\begin{equation}
    \bar{t} = \frac{1}{2^L}\sum_{i \in G} \bar{t_{i}} \,\,\,\textrm{and}\,\,\, \bar{\ell} = \frac{1}{2^L}\sum_{i \in G} \bar{\ell_{i}}\,,
\label{eq:l_and_t}
\end{equation}
where $G$ is the ensemble of all the nodes of the landscape.

To compute $\bar{t}_i$, we use the transition probabilities $P_{il}/L$ to hop from $i$ to $l$ upon a given mutation event, where $1/L$ is the probability that the mutation yields the neighbor $l$ of $i$, while $P_{il}$ is the fixation probability of this mutation, given by \cref{eq:moran} for the Moran walk or \cref{eq:transition_proba_wf} for the Wright-Fisher walk. Discriminating over all possibilities upon the first mutation, including cases where it does not fix in addition to cases where it fixes, yields
\begin{equation}
  \bar{t}_i =
    \begin{cases}
      0 & \text{if } i \in M\,,\\
      1 + \sum_{l \in G_i} \frac{1}{L} P_{il} \,\bar{t}_l + \sum_{l \in G_i} \frac{1}{L} (1-P_{il}) \,\bar{t}_i& \text{otherwise}\,,
    \end{cases}   
\label{eq:systemfsa_t}
\end{equation}
where $G_i$ is the set of neighbors of $i$ (i.e.\ the $L$ genotypes that differ from $i$ by only one mutation). 

The exact same approach can be employed to compute $\bar{\ell}_i$, but considering the normalized transition probabilities $\tilde{P}_{il}$ satisfying $\sum_{l\neq i} \tilde{P}_{il}=1$ for all $i$, instead of the raw transition probabilities $P_{il}/L$. The Markov chain associated to these normalized transition probabilities is referred to as the embedded version of the initial Markov chain. Then, we have
\begin{equation}
  \bar{\ell}_i =
    \begin{cases}
      0 & \text{if } i \in M\,,\\
      1 + \sum_{l \in G_i} \tilde{P}_{il} \,\bar{\ell}_l & \text{otherwise}\,.
    \end{cases}   
\label{eq:systemfsa_l}
\end{equation}

Solving the system of $2^L$ equations in \cref{eq:systemfsa_t} (resp. \cref{eq:systemfsa_l}) yields $\bar{t_i}$ (resp. $\bar{\ell_i}$) for all $i$, which then allows us to compute $\bar{t}$ (resp. $\bar{\ell}$) using \cref{eq:l_and_t}.

\section{Adaptive walk models}
\label{sec:AW_models}

Adaptive walks (AWs) are walks in genotype space where deleterious mutations cannot fix. Hence, the population only goes uphill in fitness until it reaches a peak, which is an absorbing state~\cite{orr2005}. Here, we present a reminder of some AW models, see also~\cite{nowak2015}. These AWs are used as references in \cref{fig:average_L=3_K=1} to compare with the Moran walk and the Wright-Fisher walk.

\paragraph{Natural AW.} At each step, the transition probability from the current wild-type genotype $i$ to a neighboring genotype $j$ is 0 if the fitness of $j$ is smaller than the fitness of $i$ ($f_j < f_i$). Conversely, if $f_j>f_i$, it reads
\begin{equation}
    P_{ij} = \frac{f_j - f_i}{\sum_{k \in B_i} (f_k - f_i)}\,,
\end{equation}
where $B_i$ is the set of neighbors of $i$ that have a larger fitness than $i$~\cite{gillespie1983,Gillespie84,Neidhart11}. Note that there are known analytical results on the natural AW under specific hypotheses, see e.g.~\cite{Neidhart11, jain2011number}.

\paragraph{Random AW.} At each step, the next genotype is chosen uniformly at random among the fitter neighbors of the current wild-type genotype~\cite{macken1989}.

\paragraph{Greedy AW.} At each step, the next genotype is the fittest among the fitter neighbors of the current wild-type genotype~\cite{Kauffman1987}.

\paragraph{Reluctant AW.} At each step, the next genotype is the least fit among the fitter neighbors of the current wild-type genotype~\cite{parisi2003, nowak2015}.

\section{Fitness landscape models considered}
\label{sec:landscape_models}

\begin{figure}[h!]
 \centering
 \includegraphics[width=\textwidth]{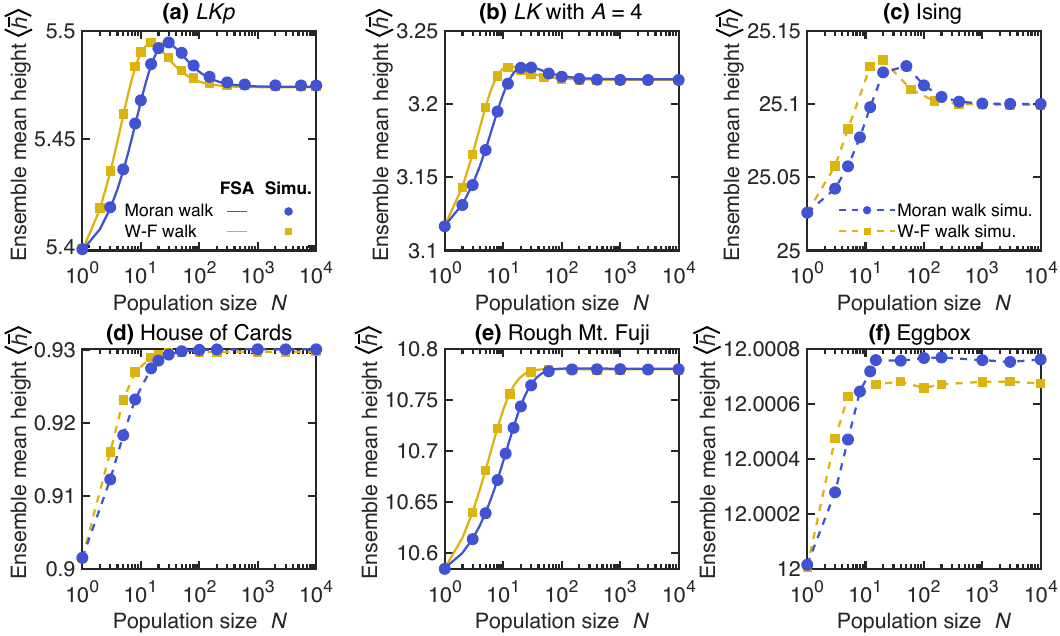}
 \caption{\textbf{Population size dependence of early adaptation in various model fitness landscape ensembles.} In each panel, the ensemble mean height $\left\langle \bar{h} \right\rangle$ of the first fitness peak reached when starting from a uniformly chosen initial node is plotted versus population size $N$, for the Moran and Wright-Fisher (W-F) walks.  All model definitions, and associated references, are given in the Supplementary material, \cref{sec:landscape_models}. (a) $LKp$ landscapes with $L = 8$, $K = 1$, and a probability $p = 0.05$ that a fitness contribution has value $q = 0.3$ instead of being drawn from a uniform distribution between 0 and 1. (b) $LK$ landscapes with alphabet size $A=4$ (i.e.\ 4 possible states at each locus), $L=4$ and $K=1$. (c) Ising landscapes with $L=8$; the couplings $J_{i,i+1}$ are drawn from a standard normal distribution and a constant offset $B = 20$ is added to the fitness of all genotypes. (d) House of Cards landscapes with $L=8$ where all genotype fitnesses are independent and drawn from a uniform distribution between 0 and 1. (e) Rough Mount Fuji landscapes with $L = 5$, reference fitness $f'_0 = 10$, additive effect $C = 0.3$ of each mutation, and epistatic contributions drawn from a standard normal distribution. (f) Eggbox landscapes with $L = 10$ and where fitnesses are drawn from two Gaussian distributions with means $f_0 \pm \mu_E/2$ and standard deviation $0.1$, with $f_0 = 10$ and $\mu_E = 4$. In all panels, markers are simulation results averaged over at least 1 walk per starting node in each landscape. In (a), (b) and (e), solid lines correspond to numerical resolutions of the FSA equations for each landscape, while markers are linked with dashed lines in (c), (d) and (f). The average denoted by $\left\langle.\right\rangle$ is performed over $10^5$ (a-d) and $10^4$ (e-f) landscapes of the ensemble considered.}
 \label{fig:overview_models}
\end{figure}

\begin{figure}[h!]
 \centering
 \includegraphics[width=0.5\textwidth]{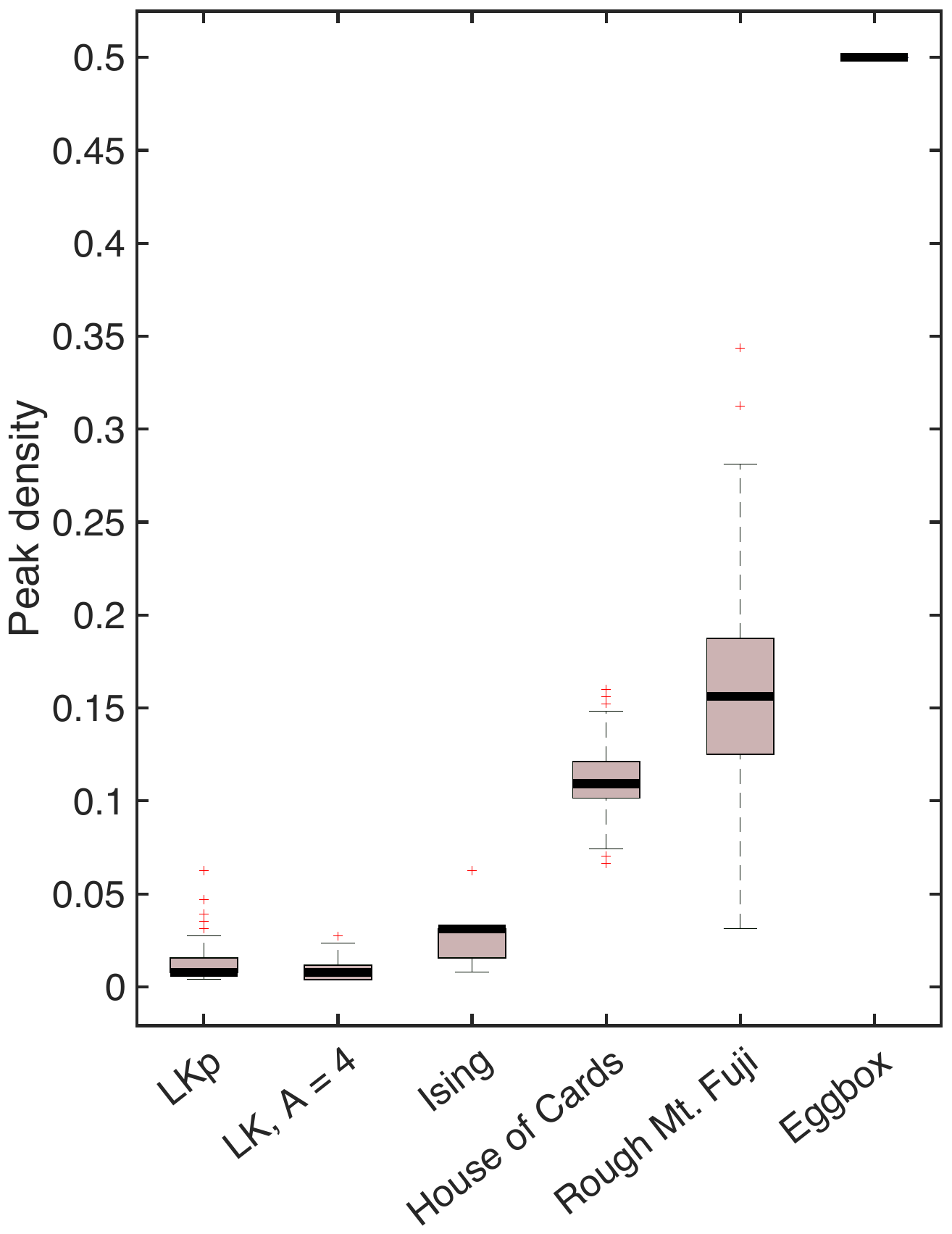}
 \caption{\textbf{Peak density in various fitness landscape models.} The distribution of the peak density (i.e., the number of peaks divided by the total number $2^L$ of genotypes) is shown as box plots for the model fitness landscape ensembles considered in \cref{fig:overview_models}. For plateaus (in the $LKp$ model), each genotype on the plateau is counted as a separate peak, yielding a generous estimate of the peak density. Bold black line: median; colored boxes: 25th and 75th percentiles; dashed lines: minimum and maximum values that are not outliers; red crosses: outliers. For each model, peaks were counted in $10^4$ fitness landscapes. The values of the parameters within each model are given in the caption of \cref{fig:overview_models}. }
\label{fig:peak_density_models}
\end{figure}

Apart from the $LK$ model described and used thoroughly in \cref{subsec:LK_model}, we consider several other models in \cref{subsec:overview}, esp. in \cref{fig:overview_models}. Here, we briefly present each of them.

\paragraph{$LKp$ landscapes.} The $LKp$ model~\cite{barnett1998} is a variant of the $LK$ model in \cref{eq:$LK$_model} where the fitness contribution $f_i$ of a given combination of states $\{\sigma_j\}_{j\in\nu_i}$ has a probability $p$ to be equal to 0 instead of being drawn from a uniform distribution between 0 and 1. It coincides with the $LK$ model for $p=0$, and becomes a completely flat landscape if $p=1$. Hence, the larger $p$ and the smaller $K$, the more likely a mutation is to be neutral. Because fitnesses equal to 0 are problematic in the Moran walk, we consider a variant of the $LKp$ landscape, where there is a probability $p$ that a fitness contribution is equal to $q>0$ (instead of 0). Note that the presence of neutral mutations implies possible fitness plateaus in these landscapes, and we consider them as peaks, meaning that the walk is stopped once the first plateau is reached.

\paragraph{$LK$ landscapes with alphabet size $A > 2$.} Here, $A$ is the number of possible states of each genetic unit, which is 2 in the usual $LK$ model. Thus, this model is a variant of the $LK$ model in \cref{eq:$LK$_model} where each node has $AL$ neighbors instead of $L$~\cite{zagorski2016}. Note that genotype space is not a hypercube in this case. Apart from this, everything is the same as in the $LK$ model considered elsewhere in this paper, in particular fitness contributions are drawn from a uniform distribution between 0 and 1. 

\paragraph{Ising landscapes.} In this model, fitness is written as (minus) the Hamiltonian of a one-dimensional Ising spin chain where each site only interacts with its closest neighbors along the chain~\cite{diu1989,ferretti2016}:
\begin{equation}
    f(\vec{\sigma}) = B + \sum_{i=1}^{L-1} J_{i,i+1}s_is_{i+1}\,,
\end{equation}
where $B$ is a positive constant that we add to avoid negative fitness values, while the $J_{i,i+1}$ are drawn from a Gaussian of fixed mean and standard deviation, and for all $i$, $s_i = 2\sigma_i-1$ so that $s_i \in \{-1, 1\}$.

\paragraph{House of Cards landscapes.} The House of Cards model is a benchmark for high ruggedness~\cite{Kauffman1987}, and its name was introduced in~\cite{kingman1978}. It is the simplest rugged fitness landscape model, because all fitness values are independent and identically distributed. Here, to generate House of Cards landscapes (\cref{fig:overview_models}(d)), we draw fitnesses from a uniform distribution between 0 and 1. Note that $LK$ landscapes with $K = L-1$ are House of Cards landscapes. This corresponds to all sites interacting together, and it yields a completely uncorrelated fitness landscape.

\paragraph{Eggbox landscapes.} In an eggbox landscape~\cite{ferretti2016}, half of the genotypes are local maxima. The high fitnesses are drawn from a Gaussian of mean $f_0 + \mu_E/2$ while the low fitnesses are drawn from a Gaussian of mean $f_0 - \mu_E/2$. Both Gaussian distributions have the same standard deviation, chosen small compared to $\mu_E$ to ensure that all non-maxima are surrounded by maxima and vice-versa.

\paragraph{Rough Mount Fuji landscapes.} In the Rough Mount Fuji model~\cite{aita2000,szendro2013b}, fitnesses have an additive part and an epistatic one:
\begin{equation}
    f(\vec{\sigma}) = f'_0 - C\,d(\vec{\sigma}_0, \vec{\sigma})+ \eta(\vec{\sigma})\,,
\end{equation}
where $f'_0$ is the fitness of a reference genotype $\vec{\sigma}_0$, while $C$ is the additive fitness effect of any mutation~\cite{szendro2013b} (note that in the original model of~\cite{aita2000}, mutations can have different additive effects), and $d(\vec{\sigma}_0, \vec{\sigma})$ denotes the Hamming distance between $\vec{\sigma}_0$ and $\vec{\sigma}$. Finally, $\eta(\vec{\sigma})$ corresponds to the epistatic contribution to fitness, and is drawn for each $\vec{\sigma}$ from a Gaussian of fixed mean and standard deviation.

\paragraph{Tradeoff-induced landscapes.} This landscape family aims to model the impact of antibiotic resistance mutations in bacteria, in particular the fact that mutations that increase the fitness of bacteria at high antibiotic concentration often decrease their fitness in the absence of antibiotic~\cite{das2020,dasPreprint}. Fitnesses are given by:
\begin{equation}
    f(\vec{\sigma},c) =\frac{r(\vec{\sigma})}{1+\left(\frac{c}{m(\vec{\sigma})}\right)^a}\,,
\end{equation}
where $c$ is the antibiotic concentration and we take $a=2$ (this exponent is typically 2 or 4 in~\cite{das2020,dasPreprint}), while $r(\vec{\sigma}) = \Pi_{i=1}^L r_i^{\sigma_i}$ and $m(\vec{\sigma}) = \Pi_{i=1}^L m_i^{\sigma_i}$. For the wild type $\vec{0}=(0,\dots,0)$, $r(\vec{0}) = 1$ and $m(\vec{0}) = 1$. The single mutant at site $i$ from the wild type is described by its fitness $r_i<1$ at $c=0$ and its resistance value $m_i>1$, and the effects of different mutations (i.e.\ the $r_i$ and $m_i$) are assumed to be multiplicative, yielding $r(\vec{\sigma})$ and $m(\vec{\sigma})$. The parameters $r_i$, $m_i$ of single mutations are independently drawn from a joint probability density $P(r_i,m_i)$ given by Eq.~(8) of~\cite{das2020}, namely:
\begin{equation}
    P(r_i, m_i) = 6\,r_i\left(1-r_i\right)\left(m_i-\frac{1}{\sqrt{r_i}}\right)e^{-\left(m_i-\frac{1}{\sqrt{r_i}}\right)}\,.
\end{equation}
\newpage

\section{Supplementary figures}

\begin{figure}[h!]
 \centering
 \includegraphics[width=0.9\textwidth]{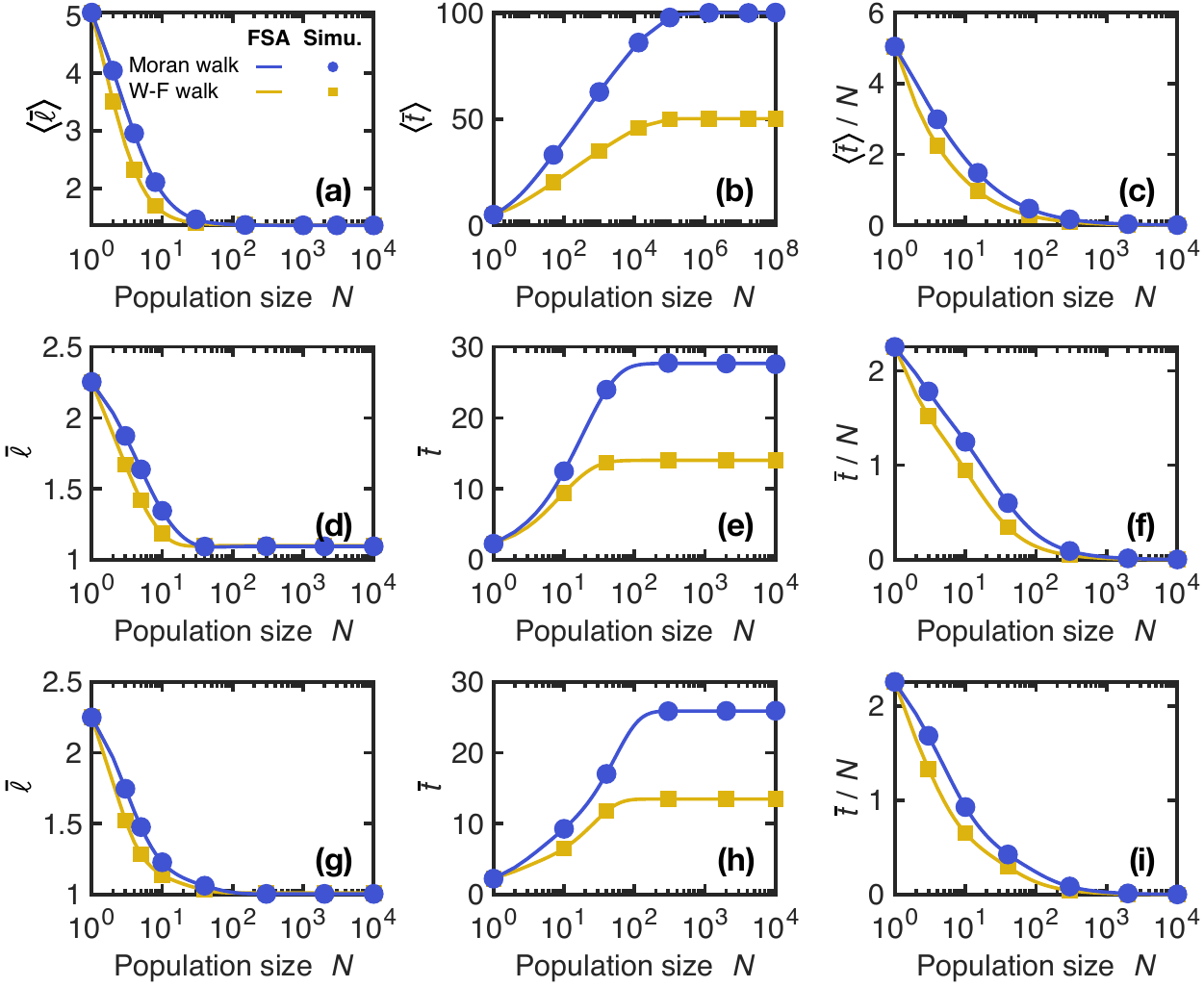}
 \caption{\textbf{Impact of population size on early adaptation in $LK$ landscapes with $L=3$ and $K=1$: time to reach the first peak.} 
 (a) Ensemble mean length $\left\langle \bar{\ell} \right\rangle$ of the walk, i.e.\ number of mutation fixation events until the first fitness peak is reached, when starting from a uniformly chosen initial node, versus population size $N$. Lines correspond to numerical resolutions of the FSA equations for each landscape, while markers are simulation results averaged over $100$ walks per starting node in each landscape (yielding $\bar{\ell}$ for each landscape). In both cases, the ensemble average denoted by $\left\langle.\right\rangle$ is performed over $5.6\times 10^5$ $LK$ landscapes with $L=3$ and $K=1$. (b) Ensemble mean time $\left\langle \bar{t} \right\rangle$ of the walk, i.e.\ number of mutation events (fixing or not) until the first fitness peak is reached, when starting from a uniformly chosen initial node, versus $N$. Symbols are as in (a); simulation results are averaged over $5\times 10^3$ walks per starting node in each landscape and ensemble averages are performed over $ 10^4$ $LK$ landscapes with $L=3$ and $K=1$. (c) Same data as (b) but time is normalized by population size $N$ to account for the fact that mutation rate is proportional to $N$. (d) Mean length $ \bar{\ell} $ of the walk, when starting from a uniformly chosen initial node, versus $N$, in landscape A, studied in \cref{fig:average_L=3_K=1}(c). Lines correspond to numerical resolutions of the FSA equations, while markers are simulation results averaged over $10^5$ walks per starting node. (e) Mean time $\bar{t}$ of the walk, when starting from a uniformly chosen initial node, versus $N$, in landscape A. (f) Same data as (e) but time is normalized by $N$. (g-i) Same as in (d-f), but in landscape B, studied in \cref{fig:average_L=3_K=1}(d), with simulation results averaged over $5\times10^5$ walks per starting node.}
 \label{fig:t_and_l_vs_N}
\end{figure}

\begin{figure}[h!]
 \centering
 \includegraphics[width=\textwidth]{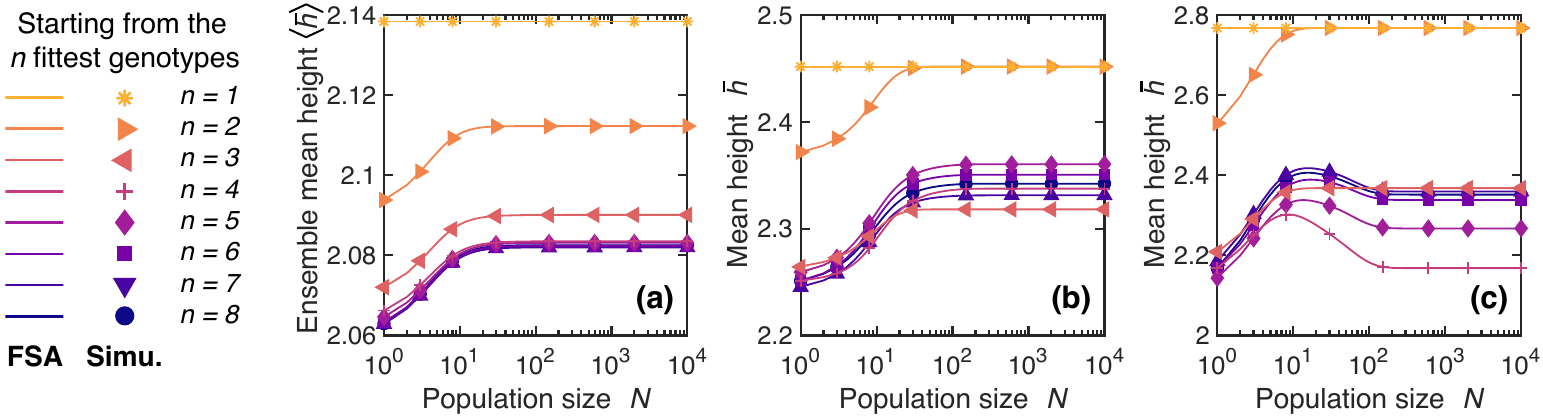}
 \caption{\textbf{Impact of population size on early adaptation in $LK$ landscapes with $L=3$ and $K=1$ for various sets of starting genotypes.} (a) Ensemble mean height $\left\langle \bar{h} \right\rangle$ of the first fitness peak reached versus population size $N$ for the Moran walk, when starting from a set of initial genotypes comprising the $n$ fittest genotypes of the landscapes, for $n$ between 1 and $2^3=8$. The starting genotype is uniformly chosen within this set. Lines correspond to numerical resolutions of the FSA equations for each landscape in a set of $2\times10^5$ landscapes with $L=3$ and $K=1$, while markers are simulation results averaged over $100$ walks per starting node in each landscape (yielding $\bar{h}$ for each landscape). (b-c) Average height $\bar{h}$ versus $N$ in the landscapes A and B considered in \cref{fig:average_L=3_K=1}(c-d) (see Table~\ref{table:landscapes}) for the Moran walk. Same symbols as in (a); simulation results averaged over $10^5$ walks per starting point.}
 \label{fig:h_vs_N_varying_starting_set}
\end{figure}

\begin{figure}[h!]
 \centering
 \includegraphics[width=0.6\textwidth]{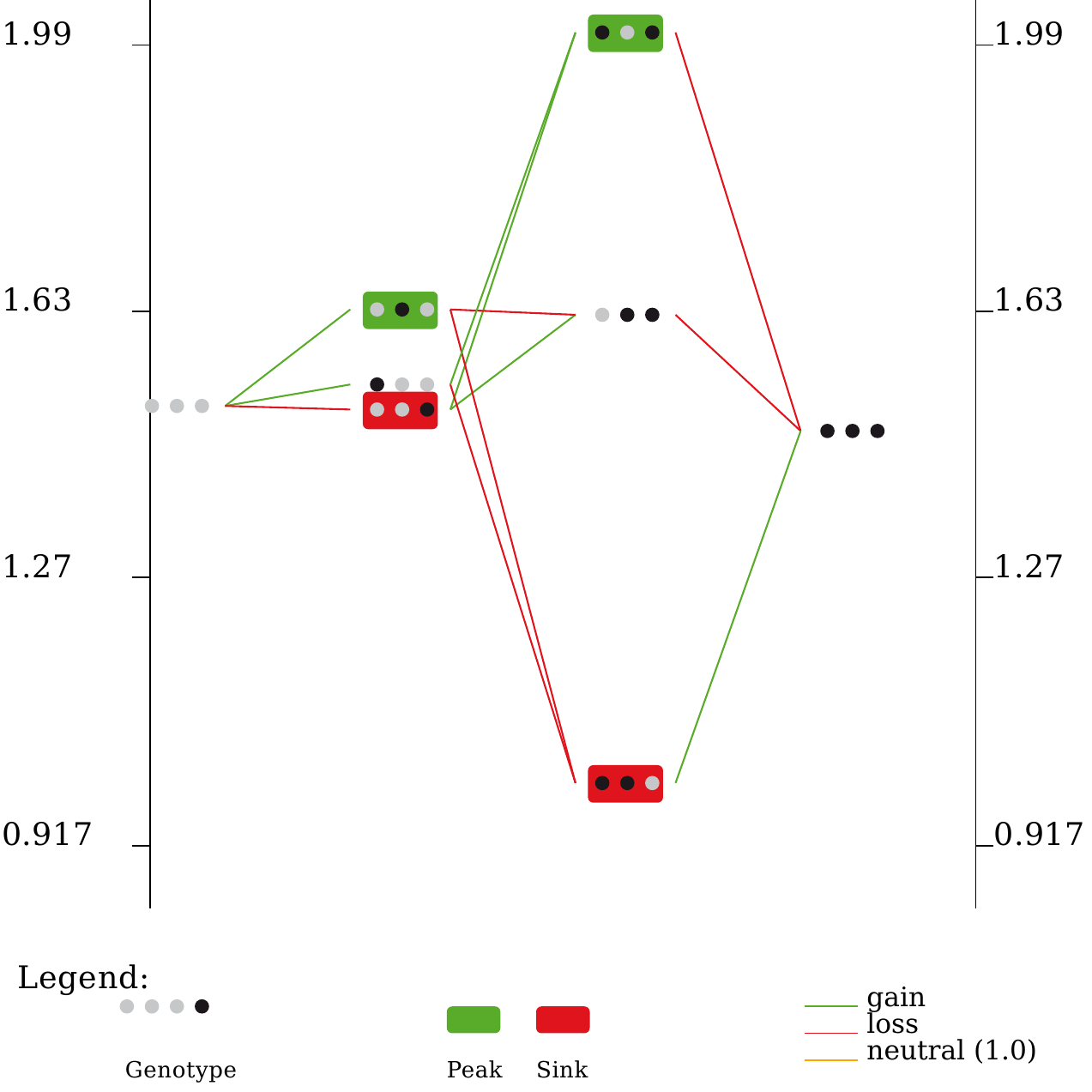}
 \caption{\textbf{MAGELLAN representation~\cite{Brouillet2015} of the fitness landscape considered in \cref{fig:accessibility_measures}(c-e).} The vertical axis denotes fitness, while the horizontal axis corresponds to Hamming distance to the wild-type. At each site, a gray marker denotes the wild-type amino acid, while a black one denotes a mutated amino acid.}
 \label{fig:landscape_fig2_magellan}
\end{figure}

\begin{figure}[h!]
 \centering
 \includegraphics[width=0.85\textwidth]{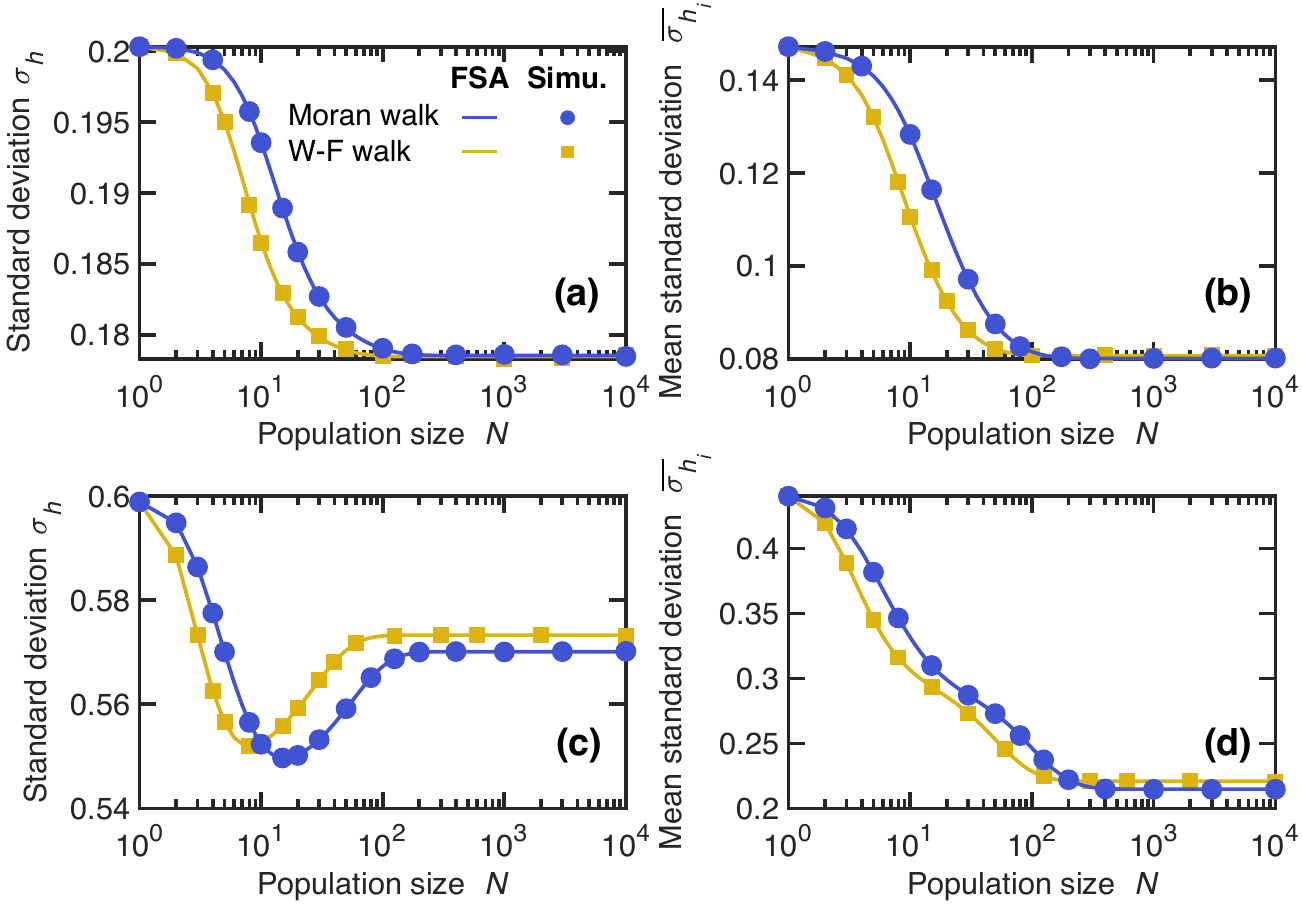}
 \caption{\textbf{Impact of population size on early adaptation in $LK$ landscapes with $L=3$ and $K=1$: standard deviation of $h$.} (a) Standard deviation $\sigma_h$ of the height $h$ of the first fitness peak reached when starting from a uniformly chosen initial node versus population size $N$. The same $LK$ landscape with $L=3$ and $K=1$ as in \cref{fig:average_L=3_K=1}(c) is considered. (b) Average $\overline{\sigma_{h_i}}$ over starting nodes $i$ of the standard deviation of $h_i$ versus population size $N$. For each starting node $i$, the standard deviation of the height $h_i$ of the first peak reached is computed, and then it is averaged over all nodes $i$ taken uniformly, yielding the average standard deviation $\overline{\sigma_{h_i}}$. Same landscape as in panel (a) and in \cref{fig:average_L=3_K=1}(c). (c) and (d) show the same as (a) and (b) respectively, but for the $LK$ landscape with $L=3$ and $K=1$ considered in \cref{fig:average_L=3_K=1}(d). In all panels, lines correspond to numerical resolution of the FSA equations and markers to simulation results obtained over (a-b) $10^5$ and (c-d) $5\times10^5$ walks per starting node.}
 \label{fig:sd_h}
\end{figure}

\begin{figure}[h!]
 \centering
 \includegraphics[width=0.45\textwidth]{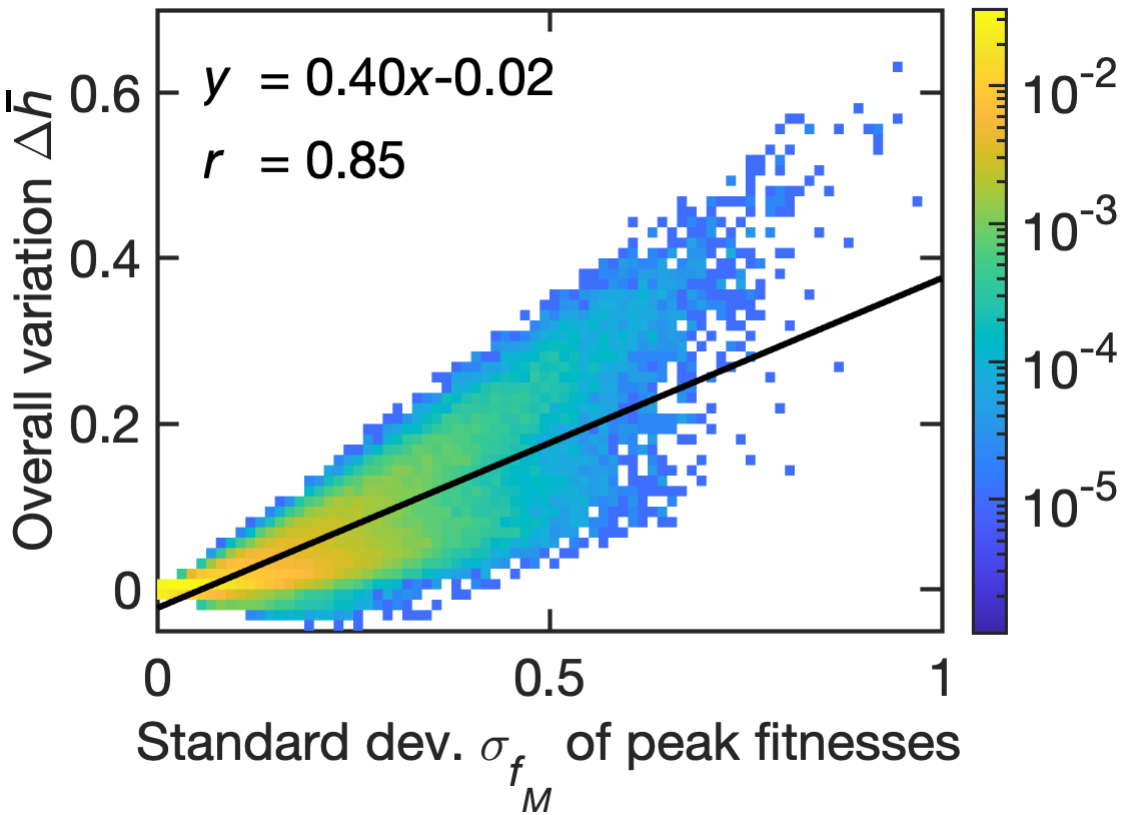}
 \caption{\textbf{Impact of the standard deviation of peak fitness values on early adaption in $LK$ landscapes.} Overall variation $\Delta \bar{h} = \bar{h}_\infty - \bar{h}(N = 1)$ versus the standard deviation $\sigma_{f_M}$ of the peak fitness values $f_M$, for the Moran walk on landscapes with $L = 3$ and $K = 1$. Color: frequency of the landscapes in the pixel; black line: linear fit (equation and Pearson correlation coefficient $r$ are given; similar results are obtained for the Wright-Fisher walk -- then, the linear fit has equation $y = 0.38 x-0.02$ and the Pearson correlation is $r = 0.83$). Here, $\Delta\bar{h}$ is obtained by numerical resolutions of the FSA equations in each landscape of an ensemble of $2\times10^5$ landscapes with $L = 3$ and $K = 1$. The large-$N$ limit $\bar{h}_\infty$ of $\bar{h}$ is evaluated for $N=10^4$.}
 \label{fig:sd_fitness_of_the_peaks}
\end{figure}

\begin{figure}[h!]
 \centering
 \includegraphics[width=0.8\textwidth]{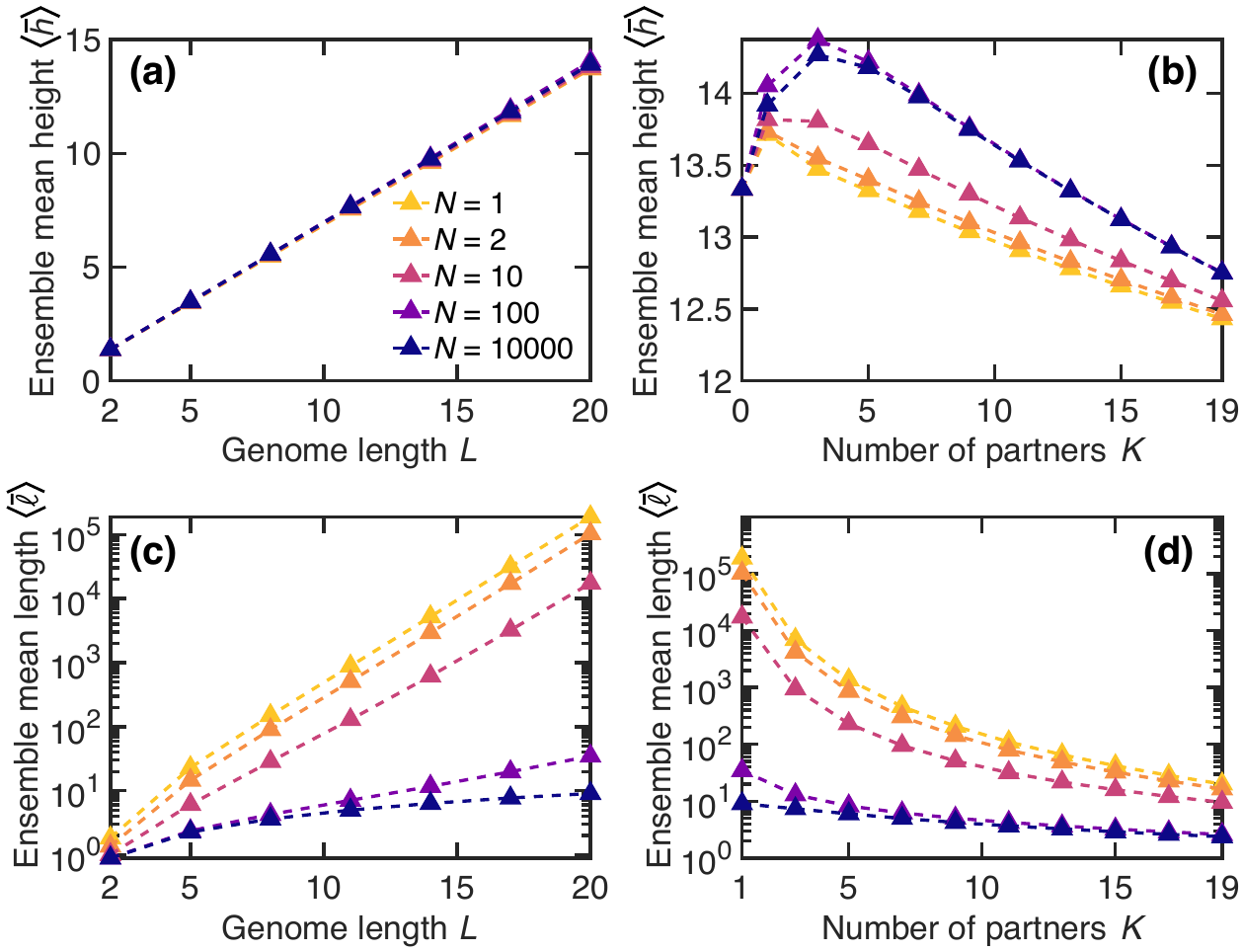}
 \caption{\textbf{Impact of $L$ and $K$ on early adaptation in $LK$ landscapes.} (a) Ensemble mean height $\left\langle \bar{h} \right\rangle$ of the first fitness peak reached when starting from a uniformly chosen initial node, versus genome length $L$, for various population sizes $N$, in $LK$ landscapes with $K=1$. (b) $\left\langle \bar{h} \right\rangle$ versus the number $K$ of epistatic partners for various $N$, in $LK$ landscapes with $L=20$. Note that the curve for $N=100$ is higher than that for $N=10^4$ at moderate values of $K$. (c) Ensemble mean length $\left\langle \bar{\ell} \right\rangle$ of the walk until the first fitness peak is reached (when starting from a uniformly chosen initial node) versus $L$ for various $N$, in $LK$ landscapes with $K=1$. (d) $\left\langle \bar{\ell} \right\rangle$ versus $K$ for various $N$, in $LK$ landscapes with $L=20$.  All results are shown for the Moran walk simulation averaged over $6\times10^5$ walks ($10^7$ walks for $L = 2$), where each walk is performed in a different landscape, generated along the way to save memory.}
 \label{fig:vsL_K}
\end{figure}

\begin{figure}[h!]
 \centering
 \includegraphics[width=\textwidth]{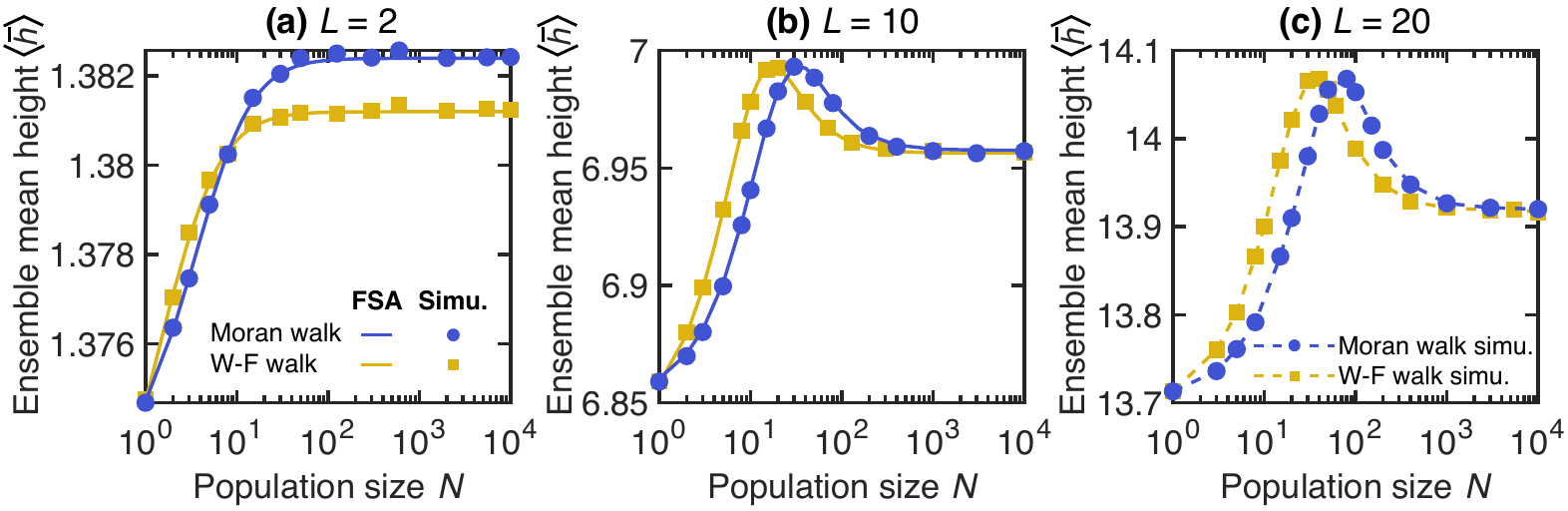}
 \caption{\textbf{Impact of population size on early adaptation in $LK$ landscapes with various genome lengths $L$.} (a) Ensemble mean height $\left\langle \bar{h} \right\rangle$ of the first fitness peak reached when starting from a uniformly chosen initial node versus population size $N$. Results are shown for the Moran and Wright-Fisher (W-F) walks for $LK$ landscapes with $L=2$ and $K=1$. (b) Same figure but for $LK$ landscapes with $L=10$ and $K=1$. (c) Same figure but for $LK$ landscapes with $L=20$ and $K=1$. In panels (a-b), lines correspond to numerical resolutions of the FSA equations for each landscape, the ensemble average being performed over $10^6$ (a) and $10^4$ (b) landscapes. In all panels, markers (linked with dashed lines in panel (c)), are simulation results averaged over (a) $10^7$, (b) $10^6$ and (c) $6\times10^5$ walks, where each walk is carried out in a different landscape, generated along the way to save memory.}
 \label{fig:h_vs_N_multiple_L}
\end{figure}

\begin{figure}[h!]
 \centering
 \includegraphics[width=0.85\textwidth]{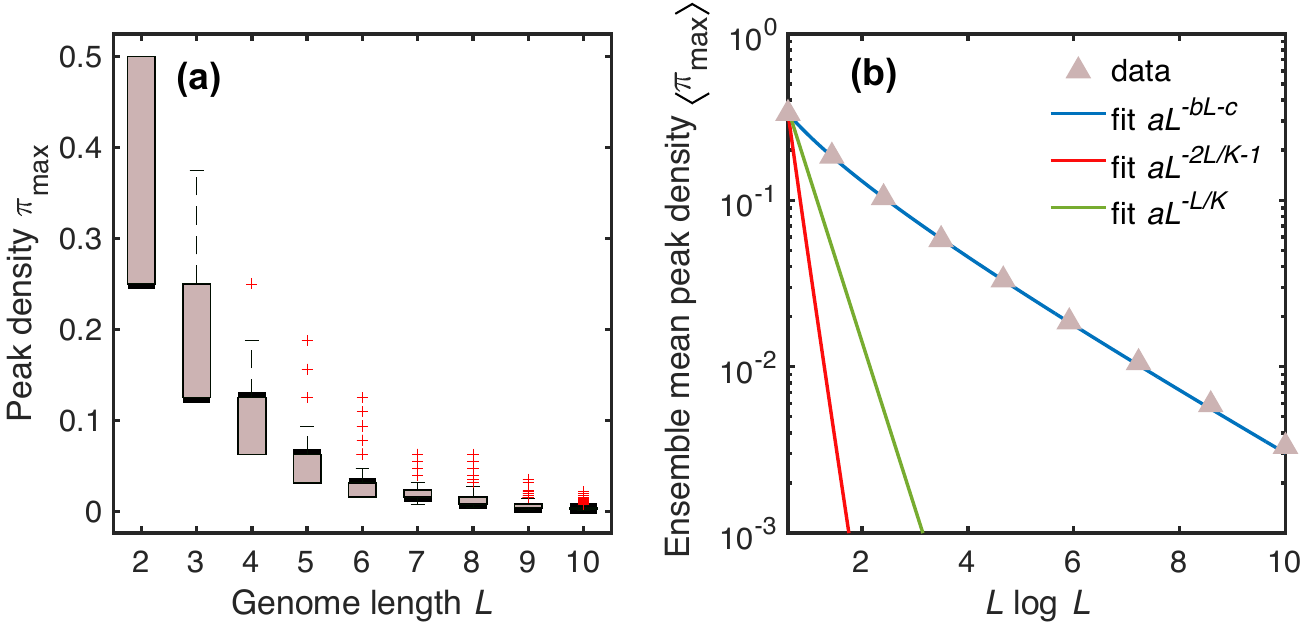}
 \caption{\textbf{Impact of genome length $L$ on peak density in $LK$ landscapes with $K = 1$.} (a)  The distribution of the peak density $\pi_\textrm{max}$ (i.e., the number of peaks divided by the total number $2^L$ of genotypes) is shown as box plots versus $L$ for $K = 1$. Bold black line: median; colored boxes: 25th and 75th percentiles; dashed lines: minimum and maximum values that are not outliers; red crosses: outliers. For each $L$, peaks were counted in $10^4$ fitness landscapes. (b) The ensemble mean peak density $\left< \pi_\textrm{max}\right>$ calculated over the same landscapes as in (a) is shown versus $L \log L$. Blue curve: fit of the form $aL^{-bL + c}$ with $a=0.671$, $b=0.167$ and $c=-0.680$; red curve: fit of the form $aL^{-2L/K - 1}$, imposing $K = 1$ (analytical prediction $\pi_{\text{max}}^{\text{MF}}$ in Eq.~(40) of~\cite{hwang2018}), where $a = 10.7$; green curve: fit of the form $aL^{-L/K}$, imposing $K = 1$ (analytical prediction $\pi_{\text{max}}^{\text{AN,BN}}$ of~\cite{hwang2018}), where $a = 1.41$. }
\label{fig:peak_density_L}
\end{figure}

\begin{figure}[h!]
 \centering
 \includegraphics[width=\textwidth]{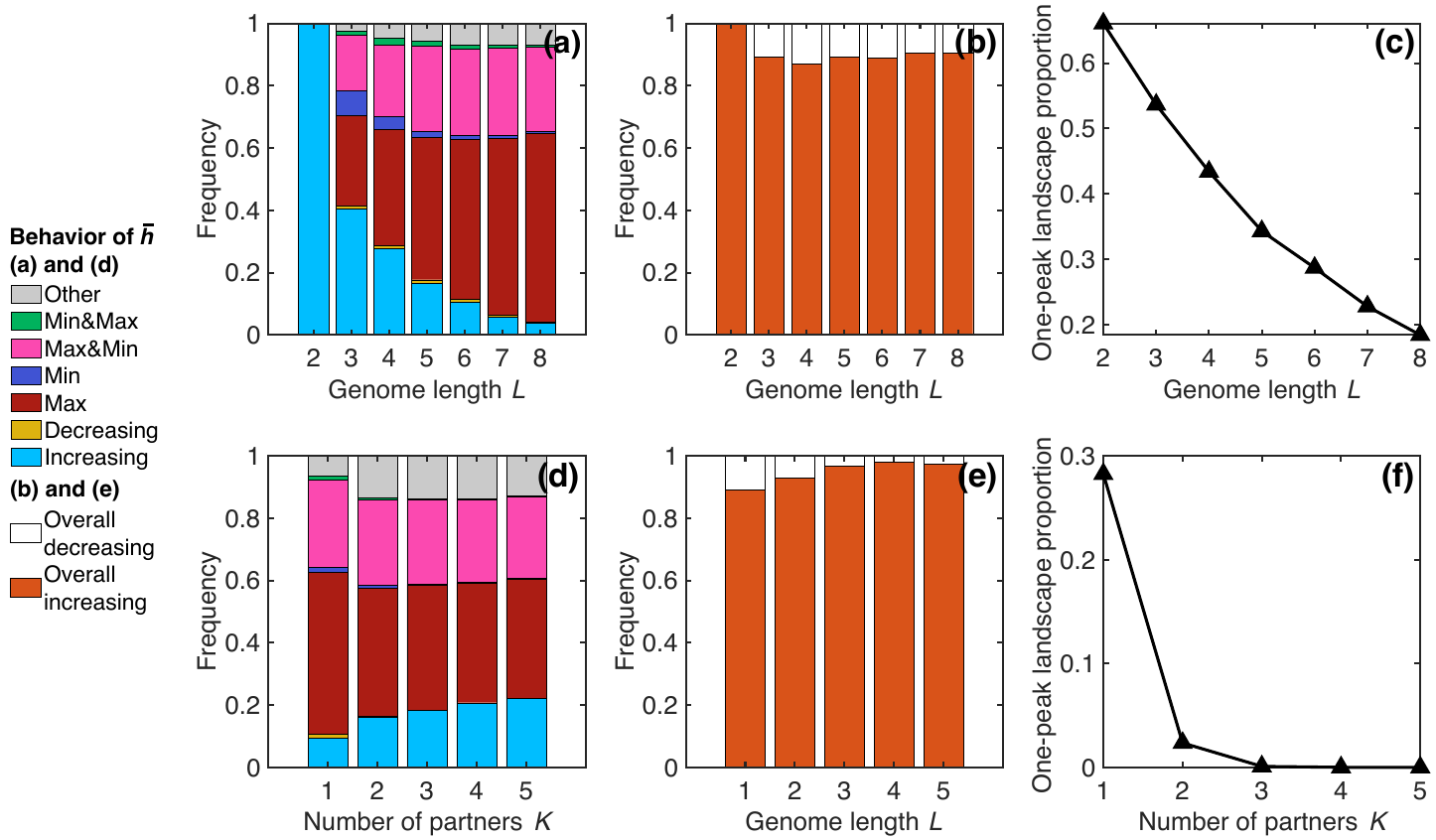}
 \caption{\textbf{Impact of $L$ and $K$ on the behavior of $\bar{h}$ in $LK$ landscapes.} (a) Frequencies of the different behaviors displayed by $\bar{h}$ versus $N$ for the Moran walk in the ensembles of $LK$ landscapes with given $L$ and $K = 1$, shown versus $L$. One-peak landscapes are discarded for this analysis. (b) Frequency of overall decreasing and overall increasing behaviors of $\bar{h}$ versus $N$ for the Moran walk in the ensembles of $LK$ landscapes with given $L$ and $K = 1$, versus $L$. (c) Proportion of one-peak landscapes in the ensembles of $LK$ landscapes with $K = 1$, versus $L$. (d-e-f) Same as (a-b-c) but versus $K$ for $L = 6$. Classes of behaviors of $\bar{h}$ versus $N$ (in (a) and (d)) are: monotonically increasing or decreasing, one maximum, one minimum, one maximum followed by a minimum at larger $N$ (``Max\&Min''), vice-versa (``Min\&Max''), and more than two extrema (``Other''), as in~\cref{fig:average_L=3_K=1}. All results are computed over $10^4$ landscapes in each class, except for $L = 3$ and $K=1$ where we use the set of $2\times 10^5$ landscapes discussed in \cref{subsec:LK_model}. In each landscape, we numerically solve the FSA equations for various $N$ to determine the behavior of $\bar{h}$.}
 \label{fig:class_proportions}
\end{figure}

\begin{figure}[h!]
 \centering
 \includegraphics[width=\textwidth]{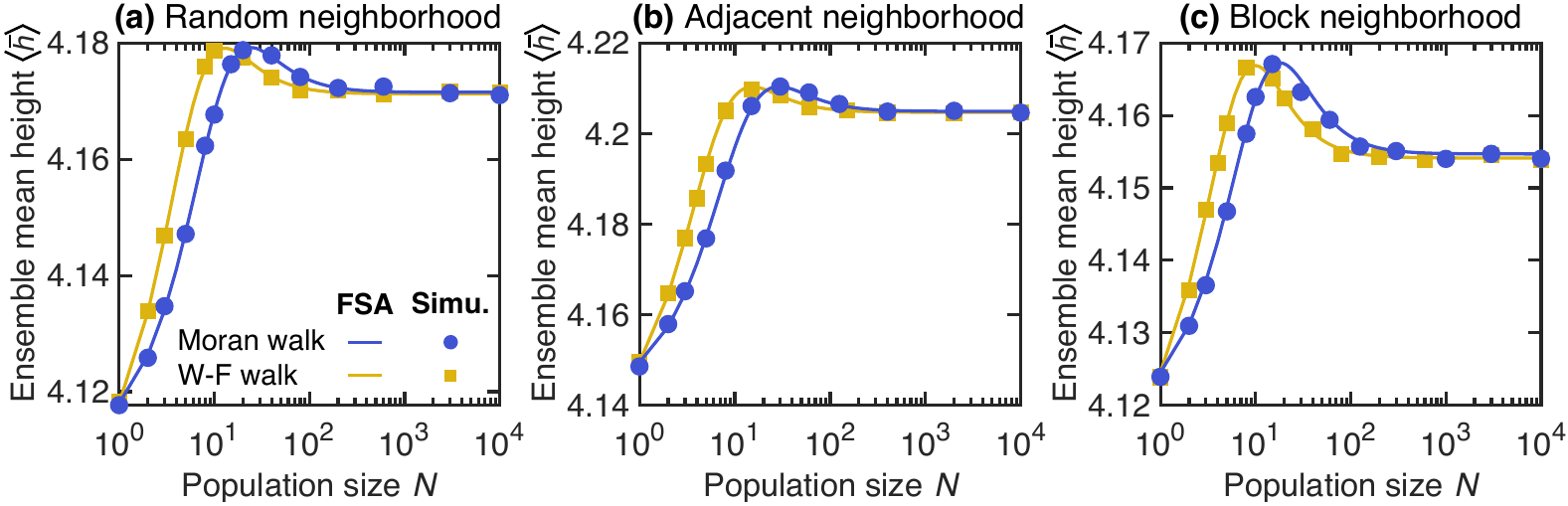}
 \caption{\textbf{Impact of population size on early adaptation in $LK$ landscapes with $L = 6$, $K = 1$ and various neighborhoods.} (a) Ensemble mean height $\left\langle \bar{h} \right\rangle$ of the first fitness peak reached when starting from a uniformly chosen initial node versus population size $N$. Results are shown for the Moran and Wright-Fisher (W-F) walks for $LK$ landscapes with random neighborhoods. (b) Same figure but for $LK$ landscapes with adjacent neighborhoods. (c) Same figure but for $LK$ landscapes with block neighborhoods. All neighborhoods types are described in~\cite{nowak2015}. In all panels, lines correspond to numerical resolutions of the FSA equations for each landscape, the ensemble average being performed over $5\times10^5$ landscapes, while markers are simulation results averaged over $10^6$ walks, where each walk is carried out in a different landscape, generated along the way to save memory.}
 \label{fig:neighborhoods}
\end{figure}

\begin{figure}[h!]
 \centering
 \includegraphics[width=\textwidth]{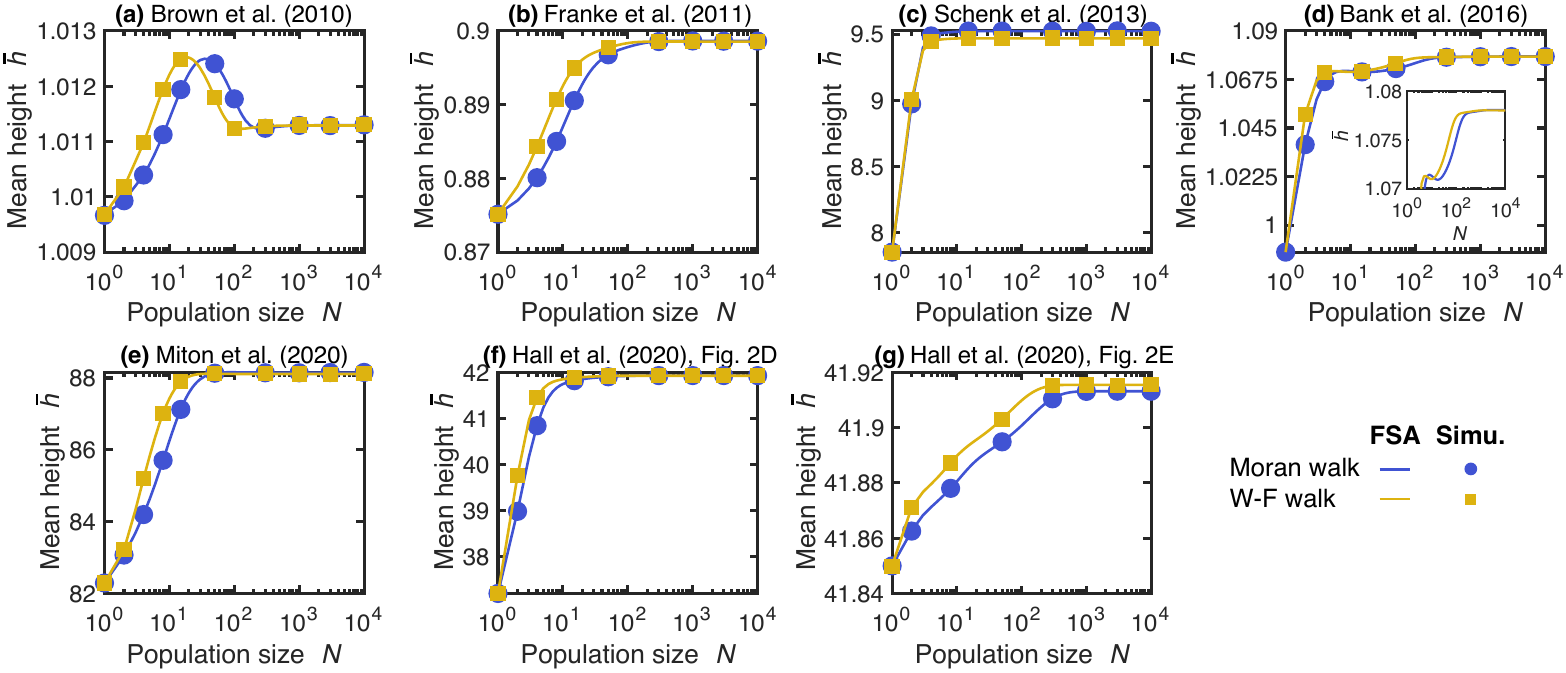}
 \caption{\textbf{Impact of population size on early adaptation in various experimental landscapes.} Mean height $\bar{h}$ of the first peak reached versus population size $N$ when starting from a uniformly chosen initial node in specific experimental landscapes. (a) Landscape from~\cite{brown2010} ($L = 5$, up to 2 different possible substitutions per site; organism: \textit{Saccharomyces cerevisiae}; fitness proxy: relative growth rate in the absence of pyrimethamine). (b) Landscape from~\cite{Franke11} ($L = 8$; organism: \textit{Aspergillus niger}; fitness proxy: (mycelium) relative growth rate; genotypes considered non-viable in~\cite{Franke11} are given a fixation probability of 0, and we do not start any walk from them). (c) Landscape from~\cite{schenk2013} ($L = 4$; organism: \textit{Escherichia coli}; fitness proxy: minimum inhibitory concentration $\text{IC}_{99.99}$ using cefotaxime). (d) Landscape from~\cite{bank2016} ($L = 6$, up to 4 different possible substitutions per site; organism: \textit{Saccharomyces cerevisiae}; fitness proxy: median growth rate; one genotype with a measured negative fitness value is given a fixation probability of 0, and we do not start any walk from it). (e) Landscape from~\cite{miton2020} ($L = 6$; organism: \textit{Escherichia coli}; fitness proxy: fold-change in esterase activity). (f-g) Landscapes from~\cite{hall2020} ($L = 7$; organism: \textit{Escherichia coli}; fitness proxy: half maximal effective concentration $\text{EC}_{50}$ using chloramphenicol). Unless otherwise specified, one substitution is considered at each site, i.e.\ sequences are binary. Here we use each fitness proxy as if it was proportional to division rate, but note that the relationship between some fitness proxies and division rate can in fact be more complex.}
 \label{fig:experimental_landscapes}
\end{figure}
\clearpage
\section{Supplementary tables}

\begin{table}[h!]
\centering
\begin{tabular}{c c c} 
 \hline
Genotype & Fitness (landscape A) & Fitness (landscape B) \\ [0.5ex] 
 \hline
 (0, 0, 0) & 1.555 & 2.199 \\  [0.5ex] 
 (0, 0, 1) & 1.674 & 0.706 \\ [0.5ex] 
 (0, 1, 0) & 2.051 & 2.767 \\ [0.5ex] 
 (0, 1, 1) & 1.181 & 1.273 \\ [0.5ex] 
 (1, 0, 0) & 2.332 & 1.524 \\ [0.5ex] 
 (1, 0, 1) & 2.452 & 1.569 \\ [0.5ex] 
 (1, 1, 0) & 2.004 & 1.481 \\ [0.5ex] 
 (1, 1, 1) & 1.134 & 1.527 \\ [0.5ex] 
 \hline
\end{tabular}
\caption{\textbf{Specific $LK$ fitness landscapes with $L=3$ and $K=1$ used in our analysis.} The first column shows genotype sequences and the next ones display the corresponding fitness values (rounded to 3 decimal places) in landscapes A and B. Landscape A is used in \cref{fig:average_L=3_K=1}(c), \cref{fig:sd_h}(a, b) and \cref{fig:t_and_l_vs_N}(d, e, f) while landscape B is used in \cref{fig:average_L=3_K=1}(d), \cref{fig:sd_h}(c, d) and in \cref{fig:t_and_l_vs_N}(g, h, i). These two landscapes were generated within the $LK$ model with $L = 3$ and $K = 1$. More precisely, epistatic partners were chosen randomly (in the random neighborhood scheme) and fitness contributions were drawn from a uniform distribution between 0 and 1.}
\label{table:landscapes}
\end{table}

\begin{table}[h!]
\centering
\begin{tabular}{c c c} 
 \hline
Site $i$ & Fitness $r_i$ & Resistance $m_i$ \\ [0.5ex]
 \hline
 1 & 0.608 & 3.399 \\  [0.5ex] 
 2 & 0.657 & 4.522 \\ [0.5ex] 
 3 & 0.630 & 1.931 \\ [0.5ex] 
 4 & 0.327 & 4.016 \\ [0.5ex] 
 5 & 0.885 & 1.443 \\ [0.5ex] 
 6 & 0.633 & 6.302 \\ [0.5ex] 
 \hline
\end{tabular}
\caption{\textbf{Fitness and resistance values used to construct the tradeoff-induced landscape studied in \cref{fig:exp_TIL}(b).} The first column shows the site indices $i$, while the corresponding fitnesses $r_i$ at antibiotic concentration $c=0$ and resistance values $m_i$ (rounded to 3 decimal places) are respectively displayed in columns 2 and 3. The values $r_i$ and $m_i$ represent the fitness at $c=0$ and the antibiotic resistance of the single mutant at site $i$ from the wild type. The tradeoff-induced fitness landscape model is presented in \cref{sec:landscape_models}.}
\label{table:r_m_TIL}
\end{table}

\end{document}